\documentclass[11pt,twoside, a4paper]{article}

\usepackage{latexsym}
\usepackage[english]{babel}
\usepackage{t1enc}
\usepackage{mathrsfs}
\usepackage{ifthen}
\usepackage{url}
\usepackage{epsfig}
\usepackage{amsbsy}
\usepackage{amsfonts}
\usepackage{amsmath}
\usepackage{amssymb}
\usepackage{amsthm}
\usepackage{amsxtra}
\usepackage{verbatim}

\numberwithin{equation}{section}
\numberwithin{figure}{section}

\newtheoremstyle{thm-style-oskari}
{7pt}      
{7pt}      
{\itshape} 
{}         
{\scshape} 
{.}        
{.5em}     
{}         

\theoremstyle{thm-style-oskari}
    \newtheorem{theorem}{Theorem}[section]
    \newtheorem{proposition}[theorem]{Proposition}
    \newtheorem{corollary}[theorem]{Corollary}
    \newtheorem{lemma}[theorem]{Lemma}
    \newtheorem{definition}[theorem]{Definition}
    
    \newtheorem{convention}{Convention}[section]
    \newtheorem{remark}{Remark}[section]

\newenvironment{Proof}[1][Proof]{\begin{proof}[\sc{#1}]}{\end{proof}}

\newcommand{\Theorem}[2] {
        \begin{theorem} \label{thr:#1}
                #2
        \end{theorem}
        }

\newcommand{\Lemma}[2] {
        \begin{lemma} \label{lmm:#1}
                #2
        \end{lemma}
        }

\newcommand{\Corollary}[2] {
        \begin{corollary} \label{crl:#1}
                #2
        \end{corollary}
        }

\newcommand{\Proposition}[2] {
        \begin{proposition} \label{prp:#1}
                #2
        \end{proposition}
        }

\newcommand{\Definition}[2] {
        \begin{definition} \label{def:#1}
                #2
        \end{definition}
        }

\newcommand{\bels}[2] {
        \begin{equation} \label{#1} \begin{split} 
                #2 
        \end{split} \end{equation}
        }
\newcommand{\bea}[1]{
	\begin{align*}
		#1
	\end{align*}
	}

\newcommand{\abs}[1]{\lvert #1 \rvert}
\newcommand{\absb}[1]{\big\lvert #1 \big\rvert}

\newcommand{\norm}[1]{\lVert #1 \rVert}
\newcommand{\normb}[1]{\big\lVert #1 \big\rVert}

\newcommand{\tE} {\mathsf{E}} 

\newcommand{\Expectation} {\tE}

\newcommand{\la} {\langle}
\newcommand{\ra} {\rangle}

\newcommand{\tP} {\mathsf{P}} 
\newcommand{\Prob} {\tP}

\newcommand{\floor} [1] {      \lfloor {#1}        \rfloor}

\newcommand{\ceil}  [1] {      \lceil  {#1}         \rceil}

\newcommand{\R} {\mathbb{R}}
\newcommand{\cR} {\bar{\R}} 
\newcommand{\C} {{\mathbb{C}}}
\newcommand{\K} {\mathbb{K}}
\newcommand{\N} {\mathbb{N}}
\newcommand{\Z} {\mathbb{Z}}

\newcommand{\T} {\mathbb{T}}

\newcommand{\Borel}{\mathcal{B}}

\newcommand{\Lspace} {\mathrm{L}} 

\newcommand{\Cont} {\mathrm{C}}

\newcommand{\Ball} {B}

\newcommand{\sett}[1] { \{ {#1} \} }

\newcommand{\setb}[1] { \bigl\{ {#1} \bigl\} }

\newcommand{\cmpl} {\mathrm{c}}

\newcommand{\titem}[1] {\item[\emph{(#1)}]} 
\newcommand{\msp}[1] {\mspace{#1 mu}}     

\newcommand{\genarg} {{\,\bullet\,}}        

\newcommand{\spt} {\mathrm{supp}}

\newcommand{\mat}[1]{\begin{bmatrix} #1 \end{bmatrix}}

\newcommand{\wti}[1] {\widetilde{#1}}

\newcommand{\dif} {\mathrm{d}}
\newcommand{\cI} {\mathrm{i}}
\newcommand{\nE} {\mathrm{e}}

\newcommand{\trans}{\mathrm{T}}

\begin{document}
\title{Rigorous scaling law for the heat current in disordered harmonic chain}
\maketitle

\begin{center}
{\Large O. Ajanki}\footnote{Partially supported by the Academy of Finland and the European Research Council}\\
Department of Mathematics, Helsinki University, \\
P.O. Box 4, 00014 Helsinki, Finland\\
\url{oskari.ajanki@iki.fi}\\
\vspace{0.5cm}

{\Large F. Huveneers}\footnote{Partially supported by the Belgian IAP program P6/02 and the Academy of Finland}\\
\vspace{0.1cm}
UCL, FYMA, 2 Chemin du Cyclotron, \\
B-1348 Louvain-la-Neuve, Belgium.\\
\url{francois.huveneers@uclouvain.be}\\
\end{center}
\vspace{2cm}

\begin{abstract}
We study the energy current in a model of heat conduction, first considered in detail by Casher and Lebowitz. 
The model consists of a one-dimensional disordered harmonic chain of $ n $ i.i.d. random masses, connected to their nearest neighbors via identical springs, and coupled at the boundaries to Langevin heat baths, with respective temperatures $ T_1 $ and $ T_n $. Let $ \Expectation \msp{1}J_n $ be the steady-state energy current across the chain, averaged over the masses. We prove that $ \tE\msp{1}J_n \sim (T_1 - T_n)\msp{1} n^{-3/2} $ in the limit $ n \to \infty $, as has been conjectured by various authors over the time. The proof relies on a new explicit representation for the elements of the product of associated transfer matrices.
\end{abstract}
\vspace{3.6cm}
\\
\emph{MSC classes:}  80A20, 82C44, 60J35

\newpage
\section{Introduction}
\label{sec:Introduction}

In a bulk of material, Fourier's law is said to hold if the flux of energy $J$ is proportional to the gradient of temperature, i.e., 
\begin{equation}\label{Fourier}
J \,=\, -\kappa\nabla T
\,,
\end{equation}
where $ \kappa $ is called the conductivity of the material.
This phenomenological law has been widely verified in practice. Nevertheless, the mathematical understanding of thermal conductivity starting from a microscopic model is still a challenging question \cite{Challange2000} \cite{Dhar-review-2008} (see also \cite{Lepri-Livi-Politi-2} for a historical perspective).

Since the work of Peierls \cite{Peierls-1929}\cite{Peierls-book-1955}, it has been understood that anharmonic interactions between atoms should play a crucial role in the derivation of Fourier's law for perfect crystals. 
It has been known for a long time that the conductivity of perfect harmonic crystals is infinite. Indeed, in this case, phonons travel ballistically without any interaction. This yields a wave like transport of energy across the system, which is qualitatively different than the diffusion predicted by the Fourier law \eqref{Fourier}. For example, in \cite{Rieder-Lebowitz-Lieb-1967}, it is shown that the energy current in a one-dimensional perfect harmonic crystal, connected at each end to heat baths, is proportional to the difference of temperature between these baths, and not to the temperature gradient. 

In addition to the non-linear interactions, also the presence of impurities causes scattering of phonons and may therefore strongly affect the thermal conductivity of the crystal. 
Thus, while avoiding formidable technical difficulties associated to anharmonic potentials, by studying disordered harmonic systems one can learn about the role of disorder in the heat conduction.
Moreover, many problems arising with harmonic systems can be stated in terms of random matrix theory, or can be reinterpreted in the context of disordered quantum systems.

Indeed, in \cite{Dhar_Spect_Dep-01} Dhar considered a one-dimensional harmonic chain of $ n $ oscillators connected to their nearest neighbors via identical springs and coupled at the boundaries to the rather general heat baths parametrized by a function $ \mu: \R \to \C $ and the temperatures $ T_1 $ and $ T_n $ of the left and right baths, respectively. Dhar expressed the steady state heat current $ J^{(\mu)}_n $ as the integral over oscillation frequency $ w $ of the modes:
\bels{intro:general stationary current as w-integral}{
J^{(\mu)}_n 
\;=\; 
(T_1-T_n) 
\int_\R 
\absb{
v_{\mu,n}^\trans\msp{-1}(w) A_n(w)\cdots A_1(w)v_{\mu,1}\msp{-1}(w) 
}^{-2} 
\dif w
\,.
}
Here $ A_k(w) \in \R^{2 \times 2} $ is the random transfer matrix corresponding the mass of the $ k $\textsuperscript{th} oscillator, while $ v_{\mu,1}(w) $ and  $ v_{\mu,n}(w) $ are $\C^2$-vectors determined by the bath function $ \mu $ and the masses of the left and the right most oscillators, respectively.   
Standard multiplicative ergodic theory \cite{RDS-Arnold-1998} tells that asymptotically the norm of $ Q_n(w) := A_n(w)\cdots A_1(w) $ grows almost surely like $ \nE^{\gamma(w)n} $ where the non-random function $ \gamma(w) \ge 0 $ is the associated Lyapunov exponent. In the context of heat conduction this corresponds the localization of the eigenmodes of one-dimensional chains while in disordered quantum systems one speaks about the one-dimensional Anderson localization \cite{Anderson-58}. 

However, in the absence of an external potential (pinning), the Lyapunov exponent scales like $ w^2 $, when $ w $ approaches zero, and this makes the scaling behavior of \eqref{intro:general stationary current as w-integral} non-trivial as well as highly dependent on the properties of the bath. 
Indeed, only those modes for which the localization length $ 1/\gamma(w) $ is of equal or higher order than the length of the chain, $ n $, do have a non-exponentially vanishing contribution in \eqref{intro:general stationary current as w-integral}. Thus the heat conductance of the chain depends crucially on how the bath vectors $ v_{\mu,1}(w), v_{\mu,n}(w) $ weight the critical frequency range $ w^2n \lesssim  1 $. 
In other words, explaining the scaling of the heat current in disordered harmonic chains reduces to understanding the limiting behavior of the matrix product $ Q_n(w) $ when $ w \leq n^{-1/2+\epsilon} $ for some $ \epsilon > 0 $.  

The evolution of $ n \mapsto Q_n(w) $ reaches stationarity only when $w^2n \sim 1 $ while the components of $ Q_n(w) $  oscillate in the scale $ wn \sim 1 $ with a typical amplitude of $ w^{-1}\nE^{\gamma_0w^2n} $ as observed numerically in \cite{Dhar_Spect_Dep-01}.
Thus the challenge when working in this small frequencies regime is that the analysis does fall back neither to classical asymptotic estimates for large $ n $, nor to the estimate of the Lyapunov exponent for small $ w $. 

Of course, the difficulty of this analysis depends also on the exact form of the vectors $ u_{\mu,k} $ in \eqref{intro:general stationary current as w-integral}, i.e., on the choice of the heat baths. Besides some rather recent developments, most of the studies so far have concentrated on two particular models. 
In the first model, introduced by Rubin and Greer \cite{Rubin-Greer-71}, the heat baths themselves are semi-infinite ordered harmonic chains distributed according to Gibbs equilibrium measures of temperatures $ T_1 $ and $ T_n $, respectively. 
Rubin and Greer were able to show that $ \tE\msp{1}J^{\text{RG}}_n \gtrsim n^{-1/2} $ with $ \tE[\genarg] $ denoting the expectation over the masses. Later Verheggen \cite{Verheggen-1979} proved that $ \tE\msp{1}J^{\mathrm{RG}}_n \sim n^{-1/2} $.

In the second model the heat baths are modeled by adding stochastic Ornstein-Uhlenbeck terms to the Hamiltonian equations of the chain (see \eqref{dynamics of the chain} below). This model, first analyzed by Casher and Lebowitz \cite{Casher-Lebowitz-71} in the context of heat conduction, was conjectured by Visscher (see ref. 9 in \cite{Casher-Lebowitz-71}) to satisfy $ \tE\msp{1}J^{\mathrm{CL}}_n \sim n^{-3/2} $.
Moreover, already in \cite{Casher-Lebowitz-71} it was argued that $ \tE\msp{1}J^{\mathrm{CL}}_n \gtrsim n^{-3/2} $. However, the line of reasoning there contains an error which invalidates this lower bound (see Section \ref{sec:Proof of Theorem}), and therefore no rigorous upper nor lower bounds have been published for $ \tE\msp{1}J^{\mathrm{CL}}_n $ until now.

\subsection{Casher-Lebowitz model and results}

The Hamiltonian of the isolated one-dimensional disordered chain is 
\begin{equation}
\label{Hamiltonian}
H(q_1, \dots q_n, p_1, \dots p_n) 
\;=\; 
\sum_{k=1}^n \frac{p_k^2}{2m_k} 
\,+\, \frac{1}{2}\sum_{k=0}^n (q_{k+1}-q_k)^2
\,, 
\end{equation} 
where $ q_k \in \R $ is the displacement of the $ k $\textsuperscript{th}
mass $ m_k $ from its equilibrium position and $ p_k $ is the associated momentum.
We consider fixed boundaries, i.e., $q_0=q_{n+1}=0$.

The usual Hamilton's equations are modified at the endpoints in order to include an interaction with heat baths.
In the Casher-Lebowitz model, this interaction consists of adding white noise and a viscous friction terms to the Hamiltonian equations of $ p_1 $ and $ p_n $: Suppose $ \lambda > 0 $ is the coefficient of viscosity, let $T_1\ge T_n > 0$ be the respective temperatures of the reservoirs, and let $ W_1, W_n $ be two independent Brownian motions. 
The equations of motion for the Casher-Lebowitz chain then take the form of the stochastic differential equation
\bels{dynamics of the chain}{
\dif q_k 
\;&=\; 
\frac{\partial H}{\partial p_k} \dif t
\,,
\\
\dif p_k \;&=- \frac{\partial H}{\partial q_k} \dif t \,+\, (\delta_{k,1}+\delta_{k,n})(-\lambda p_k \dif t + \sqrt{2\lambda T_k m_k}\,\dif W_k)
\,,
}
with $1\le k\le n $.
If $ \sett{e_1, e_2} $ is the canonical basis of $ \C^2 $, then, as far as the scaling behavior goes, the choice \eqref{dynamics of the chain} of heat baths corresponds (see \cite{Casher-Lebowitz-71}, and \eqref{v_CLpm sim v_pm} below) to setting $ v_{\mu,1}(w) = \abs{w}^{-1/2} e_1 + \cI \abs{w}^{1/2}e_2 $ and $ v_{\mu,n}(w) = \abs{w}^{-1/2} e_1 - \cI \abs{w}^{1/2}e_2 $ in \eqref{intro:general stationary current as w-integral}. 
The resulting current, denoted by $ J^{\mathrm{CL}}_n(m_1,\dots,m_n) $, is then by definition the average rate at which energy is carried from the left to the right heat bath over the stationary measure of \eqref{dynamics of the chain} for fixed masses $ m_k $.  

Now, suppose that the masses are random variables $ M_k $. Our main result is the following strict scaling relation for the mass averaged stationary current.
%
\Theorem{the scaling of the average current}{
Assume that the masses $ (M_k:k\in\N) $ are independent and identically distributed. 
Suppose that the common probability distribution of the masses $ M_k $ admits a density, 
compactly supported on $\rbrack 0,\infty\lbrack$, 
continuously differentiable inside its support, 
with an uniformly bounded derivative.
Denote by $ \tE[\genarg] $ the expectation over the masses. Then there exist $K,K'>0$ such that the heat current $ J^\mathrm{CL}_n $ satisfies the relation
\bels{the result}{
K\, \frac{T_1-T_n}{n^{3/2}}
\;\leq\; 
\tE\bigl[ 
J^{\mathrm{CL}}_n\msp{-1}(M_1,\dots M_n) 
\bigr]  
\;\leq\; 
K'\, \frac{T_1-T_n}{n^{3/2}} 
\,.
}
}
The proof is based on a new representation of the matrix $ Q_n(w) $ in terms of a discrete time Markov chain on a circle. Based on this representation we obtain a good control of the joint behavior of the matrix elements of $ Q_n(w) $ for the most important regime $ w \leq n^{-1/2+\epsilon} $ where $ \epsilon > 0 $ is small. 
Moreover, together with O'Connor's decay estimates \cite{O'Connor-75} for high frequencies we have a good control of the exponential decay of $ \norm{Q_n(w)} $ whenever $ w \ge n^{-1/2+\epsilon} $.
Therefore, the possibility of generalizing Theorem \ref{thr:the scaling of the average current} to a quite large class of heat baths seems possible by extending our analysis.  
Indeed, in Subsection \ref{ssec:Other heat baths} we sketch how one can derive the scaling behavior of the stationary heat current for Dhar's modified version of the Casher-Lebowitz model as well as to prove the analogue of Theorem \ref{thr:the scaling of the average current} for the Rubin-Greer model. 

The organization of the paper is as follows. In Section \ref{sec:conventions and outline}, we present the practical expression for the current $ J^{\mathrm{CL}}_n $, after first introducing some conventions and notation to be used in the rest of the paper. In the end of Section \ref{sec:conventions and outline} our strategy to obtain Theorem \ref{thr:the scaling of the average current} is outlined.  
Sections \ref{sec:Representation of matrix elements} to \ref{sec:Potential theory} contain the three main technical results needed for the proof. 
The actual proof of Theorem \ref{thr:the scaling of the average current} is then presented in Section \ref{sec:Proof of Theorem}. 

\section{Conventions and outline of paper}
\label{sec:conventions and outline}

For the rest of this manuscript we are going to \emph{assume that the conditions of Theorem \ref{thr:the scaling of the average current} hold.} In particular, this means that the zero mean random variables
\bels{def of B_k}{
B_k \;:=\; \frac{M_k-\tE\msp{1}M_k}{\tE\msp{1}M_k}
\,,
}
are i.i.d., have a (Lebesgue) probability density $ \tau $ that satisfies $ \spt(\tau) \subset [b_-,b_+] $, and $ \tau \in \Cont^1([b_-,b_+]) $, for some constants  $ -1 < b_- < b_+ < \infty $. Here $ \Cont^k([a,b]) $ denotes a continuous function $ f: [a,b] \to \R $ such that $ \frac{\dif^j f}{\dif x^j} $ exist for $ j \leq k $, and that these derivatives are bounded and continuous on $ ]a,b[ $. 
The transfer matrices appearing in \eqref{intro:general stationary current as w-integral} are related to $ B_k $: 
\bels{def of matrix A_k}{
A_k \;\equiv\;
A_k(w) \;=\; \mat{2 - \pi^2 w^2 (1+B_k) & -1 \\ 1 & 0}
\,,
}
where the frequency variable $ w $ is related to the frequency variable $ \omega $ in \cite{Casher-Lebowitz-71} by $ \omega = \pi^{-1} (\tE\msp{1}M_k)^{1/2} w $. 
As already pointed out in the introduction, O'Connor has shown (see Theorem 6 and its proof in \cite{O'Connor-75}) that for any reasonable heat baths the frequencies above any fixed $ w_0 > 0 $ have exponentially small contribution to the total current \eqref{intro:general stationary current as w-integral} as $ n $ grows.
%
Therefore, one may \emph{consider an arbitrary small but fixed interval $ ]0,w_0] $ of frequencies $ w $ in order to prove Theorem \ref{thr:the scaling of the average current}.}   

We write $ \N = \sett{1,2,3,\dots} $, $ \N_0 = \sett{0,1,2,\dots} $ and $ \R_+ = \;]0,\infty[\, $ and $ \bar{\R} = \R \cup \sett{\infty} $ with $ \infty = \pm \infty $.
Additionally, following conventions are used frequently.

\vspace{0.2cm}

\noindent{\bf Probability:} 
Since all the randomness of the stationary state current $ J^\text{CL}_n $ originates from the random masses we define the probability space $ (\Omega,\mathcal{F},\tP) $ as the semi-infinite countable product of spaces $ ([b_-,b_+],\Borel[b_-,b_+], \tau(b) \dif b) $. Here $ \mathcal{B}(S) $ denotes the Borel $ \sigma$-algebra of the topological space $ S $. 
The filtration generated by the sequence $ B \equiv (B_k:k\in\N) $ is denoted by $ \mathbb{F} = (\mathcal{F}_k:k\in\N)  $, $ \mathcal{F}_k = \sigma(B_j:1 \leq j\leq k) \subset \mathcal{F} $.
As a convention, the names of new random variables on $ (\Omega,\mathcal{F},\tP) $ will be generally written in capital letters. A discrete time stochastic process $ (Z_n:n\in \K) $ is denoted by $ Z \equiv (Z_n) $ when index set $ \K $ is known or not relevant. Finally, we write $ \Delta Z_n = Z_n-Z_{n-1} $.

\vspace{0.2cm}
\noindent{\bf Constants and scaling:} 
Because we are interested only in the scaling relations many expressions can be made more manageable by using the following conventions.
First, we use letters $ C, C', C_1, C_2,\dots $ to denote strictly positive finite constants, whose value may vary from place to place. 
Except otherwise stated, these values depend only on $ \tau, \lambda, T_1 - T_n $ and $ w_0 $, but never on $ w $ or $ n $.
Secondly, suppose $ f, g,h $ are functions, we write $ f \lesssim g $, or equivalently, $ g \gtrsim f $ provided $ f \leq C\, g $ pointwise, i.e., $ f(x,y) \leq C g(y,z) $ for all possible arguments $ x,y,z $.  
If $ f \lesssim g $ and $ f \gtrsim g $ then we write $ f \sim g $.
Moreover, the expression $ f = g + \mathcal{O}(h) $, where $ f,g,h $ means $ \abs{f-g} \leq C \abs{h} $. 

\vspace{0.2cm}
\noindent{\bf Periodicity:} 
In the following we are going to deal with functions that are defined and/or take values on the unit circle $ \T = \R/\Z $. The following conventions are practical on such occasions.
When $ x \in \R $ write $ \abs{x}_\T = \min(x-\floor{x},\ceil{x}-x) $ where $ \floor{x} $ ($ \ceil{x} $) denotes the largest (smallest) integer smaller (larger) than $ x $. 
We identify 1-periodic functions on $\R$ with functions on $\T$. Similarly, a function  $g:\R \to \R $ of the form $ g(x) = x + f(x) $, where $ f $ is 1-periodic, is identified with a function from $ \T $ to itself.

 
\subsection{Heat current in terms of matrix elements}

Let $ v = [v_0\msp{8}v_{-1}]^\trans \in \C^2 $, and denote by $ D(v) \equiv (D_n(v):n \in \N) $ the discrete time stochastic process that solves for $ n \in \N $:
\bels{def of process D_n(v)}{
D_n(v) \;&=\; (1-\pi^2w^2(1+B_n)) \msp{1} D_{n-1}(v) \,-\, D_{n-2}(v)
\\
D_0(v) \;&=\; v_0\,,
\\
D_{-1}\msp{-1}(v) \;&=\; v_{-1}
\,.
}
By definition one then has for $ n \in \N $
\bels{solution of Q_n in terms of D_n(v)}{
Q_n \;=\; A_n A_{n-1} \cdots A_1 \;=\; \mat{D_n(e_1)\; & D_n(e_2)\; \\ D_{n-1}(e_1) & D_{n-1}(e_2)}
\,,
}
where $ A_k $ is the transfer matrix \eqref{def of matrix A_k} and $ e_1 = [1\msp{8}0]^\trans $ and $ e_2 = [0\msp{8}1]^\trans $. 
As a remark it is worth noting that in the derivation of the stationary heat current one actually starts with \eqref{def of process D_n(v)} where $ D_n(e_k) $ are certain real valued (sub-)determinants of a semi-infinite matrix and then expresses the final formula conveniently in terms of the product \eqref{solution of Q_n in terms of D_n(v)}. 

Now, in \cite{Casher-Lebowitz-71} it was proven that Casher-Lebowitz model corresponds to setting the bath vectors $ v_{\mu,1} $ and $ v_{\mu,n} $ in the general expression \eqref{intro:general stationary current as w-integral} of $ J^{(\mu)}_n $ equal to
\bels{v_CLpm sim v_pm}{
v_{\text{CL},1}(w) \;=\; \mat{ (\alpha M_1\abs{w})^{-1/2}\,  \\ +\cI (\alpha M_1 \abs{w})^{1/2}} 
\quad\text{and}\quad
v_{\text{CL},n}(w) \;=\; \mat{ (\alpha M_n\abs{w})^{-1/2}\,  \\ -\cI (\alpha M_n \abs{w})^{1/2}}
\,.
}
Here the constant $ \alpha > 0 $ depends on the units of the frequency variable $ w $, etc. 
Since the masses have a compact support, $ [m_-,m_+] \subset \,]0,\infty[ $ and the bath vectors are symmetric in $ w $, one has
\bels{CL-current and def of J_n}{
J^{\text{CL}}_n 
\,\sim\;  
(T_1-T_n) \int_\R \abs{v_{\text{CL},n}^\trans(w) Q_n(w) v_{\text{CL},1}(w)}^{-2} \dif w 
\;\sim\; 
\int_0^\infty j_n(w) \dif w \;=:\;J_n
\,,
} 
where $ j_n(w) := \abs{v_n^\trans(w)Q_n(w)v_1(w)}^{-2} $, with $ v_1(w) = w^{-1/2}e_1 + \cI w^{1/2} e_2 $ and $ v_n(w) = w^{-1/2}e_1 - \cI w^{1/2}e_2$. 
By using $ D_n(e_1) D_{n-1}(e_2) - D_{n-1}(e_1) D_n(e_2) = \det(A_n\cdots A_1) = 1^n = 1 $ to get rid of the mixed terms of $ D_n(e_k) \equiv D_n(e_k;w) $ one obtains: 
\bels{def of jn}{
j_n(w)
\;&=\; \bigl\{1 \,+ w^{-2}D_{n}(e_1)^2 + D_{n-1}(e_1)^2 \!+ D_{n}(e_2)^2 + w^2D_{n-1}(e_2)^2\,\bigr\}^{-2}
\,.
}
This is the form we are going to use for the proof of Theorem \ref{thr:the scaling of the average current}.

\subsection{Outline of the proof}
\label{ssec:Outline}

It follows from \eqref{CL-current and def of J_n} and \eqref{def of jn} that the scaling bounds of $ \tE(J^{\text{CL}}_n) \sim \tE(J_n) $ rely on the good understanding of the processes $ D(v) $ defined in \eqref{def of process D_n(v)}. 
Thus, the first natural step towards the proof of the theorem is the derivation of an easier representation for $ D_n(v) $. This is the purpose of Section \ref{sec:Representation of matrix elements} where one constructs (Proposition \ref{prp:Fundamental decomposition of D_n(v)} and Corollary \ref{crl:Fundamental decompositions of D_1n and D_2n}) the representations: 
\bels{outline:decomposition D_1n and D_2n}{
D_n(e_1) \;\sim\; w^{-1}\Gamma_n^\vartheta \cdot \sin \pi X^\vartheta_n
\,,\qquad\text{and}\qquad
D_n(e_2) \;\sim\; w^{-1}\Gamma_n^0 \cdot \sin \pi X^0_n
\,.
}
Here $ \vartheta = w + \mathcal{O}(w^3) $ is a constant, the phases $ (X^x_n: n \in \N_0) $ form a Markov process on $ \T $ 
\bels{outline:recursion for X^x}{
X^x_n \;&=\; X^x_{n-1} +\, w \,+\, w \msp{1}\phi(X^x_{n-1})B_n \,+\, \mathcal{O}(w^2) 
\qquad\text{with}\qquad X^x_0 = x
\,,
}
and the amplitude $ \Gamma^x_n \in \,]0,\infty[ $ is an exponential functional of $ (x,B_k:1\leq k\leq n)$: 
\bels{outline: Gamma^x_n}{
\Gamma_n^x \;&=\; \nE^{w \sum_{k=1}^n  s(X^x_{k-1})B_k \;+\; w^2 \sum_{k=1}^n r(X^x_{k-1})B_k^2 \;+\; \mathcal{O}(w^3n)}
\,.
}
The smooth functions $ \phi,s,r : \T \to \R $ are explicitly known. The process $ X \equiv X^x $ is specified precisely in Definition \ref{def:Process X^x} and Lemma \ref{lmm:functions fb and Phi}, and its most important qualitative properties are listed in Corollary \ref{crl:the three qualitative properties of X}. 
The main advantage of the representation \eqref{outline:decomposition D_1n and D_2n} is that, unlike the recursion relations \eqref{def of process D_n(v)} of $ D(v)$, it allows us to treat both the scaled noise $ wB_n $ and the initial values $ e_2 $ of $ D_n(e_2) $ as small perturbations around $ 0 $ and $ e_1 $, respectively.

Based on the representation \eqref{outline:decomposition D_1n and D_2n}, let us now carry out heuristic computations which form the outline for the actual proof of $ \tE(J_n) \sim n^{-3/2} $. 
Along these calculations we will point out the properties of $ X^x_n $ and $ \Gamma^x_n $ which must be proven to make these calculations rigorous. 
We start with the upper bound. By Theorem 6 of \cite{O'Connor-75} we may restrict the integration domain of \eqref{CL-current and def of J_n} into $ [0,w_0]$. Dropping positive terms from the denominator in \eqref{def of jn} then yields 
\begin{subequations}
\begin{align}
\tE\msp{1}J^\mathrm{CL}_n 
\;\sim\;
\tE\msp{1}J_n 
\;&=\;
\tE \int_{0}^\infty  j_n(w) \dif w 
\;\leq\; 
\int_{0}^{w_0} \tE\left\{ \frac{1}{1+w^{-2}D_n(e_1;w)^2}\right\}\msp{1} \dif w
\,
\label{outline:upper bound first lines}
\\
&=\; 
\int_{0}^{w_0} \tE \left\{\int_\T\frac{1}{1+(w^{-2}\Gamma_n \sin x)^2}\,\tP(X_n \in \dif x|\Gamma_n)\right\} \dif w
\label{outline:upper bound need for pot theory} 
\,.
\end{align}
\end{subequations}
Now comes the first crucial step. By standard martingale central limit theorems \cite{Hall1980} one expects that  $ X_n $, if properly centered, scaled, and considered as a process on $ \R $, should converge to a Gaussian with unit variance. 
Unfortunately, such weak convergence results do not suffice since we need to deal with very unlikely events. 
Indeed, from \eqref{outline:upper bound need for pot theory} one sees that the crucial contribution of the terms inside the curly brackets comes when $ \abs{X_n} \lesssim w^2 / \Gamma_n $. The probability of this to happen is typically very small, e.g., of order $ n^{-1} $ when $ w^2n \sim 1 $.   
Moreover, we would also like to be able to consider $ X_n $ and $ \Gamma_n $ effectively independent in \eqref{outline:upper bound need for pot theory}. 
In other words, we would like to have:
\begin{itemize}
\item[(a)] 
Pointwise bound: 
$ 
\chi_{\Ball(wn,C w\sqrt{n})}(x) \cdot 
\frac{\dif x}{\min(1,w\sqrt{n})}
\;\lesssim\; 
\tP(X_n \in \dif x) 
\;\lesssim\; 
\frac{\dif x}{\min(1,w\sqrt{n})} $, $  x \in \T $;
\item[(b)] 
Independence: $ \tP(X_n \in \dif x|\Gamma_n) \;\sim\; \tP(X_n \in \dif x)\; $, $ x \in \T $. 
\end{itemize}
The purpose of Section \ref{sec:Potential theory} is to prove Proposition \ref{prp:Potential theory} which together with the bounds in Subsection \ref{ssec:Proof of the upper bound} implies that as far as \eqref{outline:upper bound need for pot theory} goes one may think that both (a) and (b) hold literally.
So by using (a-b) and then parametrizing $ \T $ with $ [-1/2,1/2] $ in \eqref{outline:upper bound need for pot theory} one gets
\begin{align}
\tE(J_n)\;
&\lesssim\;
\int_0^{w_0} \tE \left\{ \int_{-1/2}^{1/2} \frac{1}{1+(w^{-2}\Gamma_n x)^2}\cdot\frac{\dif x}{\min(1,w\sqrt{n})} \right\} \dif w   
\notag
\\ 
&\lesssim\;
\int_0^{w_0}
\frac{1}{\min(1,w\sqrt{n})} \tE\left\{ \frac{\arctan (w^{-2}\Gamma_n)}{w^{-2}\Gamma_n} \right\} \dif w
\notag
\\ 
&\lesssim\;
\int_0^{n^{-1/2}}\msp{-7}\frac{w}{\sqrt{n}}\,\tE\bigl\{1/\Gamma_n(w)\bigr\}\,\dif w \,+\, \int_{n^{-1/2}}^{w_0}\tE\bigl\{ 1/\Gamma_n(w)\bigr\}\, \dif w 
\label{outline:Expectation of 1/Gamma appears}
\,.
\end{align}
Here we have used the upper bound in (a), approximated $ \sin z \sim z $ and then performed a change of variables $ x \mapsto w^{-2}\Gamma_n x $. To get the last line we have approximated $ \arctan r \lesssim 1 $, for $ r \in \R_+ $. 

In Section \ref{sec:Expectation of 1/Gamma_n} we bound the only unknown term in \eqref{outline:Expectation of 1/Gamma appears} by showing that there exists a constant $ \alpha > 0 $ such that 
\begin{equation}
\label{outline:expectation of 1/Gamma}
\tE\{1/\Gamma_n(w)\} \;\lesssim\; \nE^{-\alpha w^2n}\,, 
\quad\text{when}\quad 0 < w \leq w_0
\,.
\end{equation}
The sum over $ r $-terms in \eqref{outline: Gamma^x_n} is then shown to produce an exponent $ \nE^{-\gamma(w)n} $ where the constant $ \gamma(w) \sim w^2 $ is the Lyapunov exponent associated to the transfer matrices $ A_k $ in \eqref{def of matrix A_k} with explicit value given in \eqref{Lyapunov exponent explicitly solved}. 
The challenge in Section \ref{sec:Expectation of 1/Gamma_n} is to bound the large deviations of the first sum in \eqref{outline: Gamma^x_n} so much that \eqref{outline:expectation of 1/Gamma} still holds for some $ \alpha > 0$.
By applying the bound \eqref{outline:expectation of 1/Gamma} in \eqref{outline:Expectation of 1/Gamma appears}, yields the upper bound for the total current:   
\bea{
\tE(J_n)\;
&\lesssim\;
\int_0^{n^{-1/2}}\msp{-15}\frac{w}{\sqrt{n}}\cdot 1\, \dif w 
\,+\,
\int_{n^{-1/2}}^{w_0} w^2 \nE^{-\gamma w^2 n} \dif w \;\sim\; n^{-3/2} 
\,.
}

To prove the lower bound, it suffices to show that for $ w \in I := [(2n)^{-1/2},n^{-1/2}] $ one has $ \tP\bigl(j_n(w) \ge C w^2 \bigr) \gtrsim 1 $. Indeed, if this bound is verified then 
\[
\tE(J_n) \gtrsim \int_I \tE \msp{1} j_n(w)\msp{1} \dif w 
\;\ge\;  
n^{-1/2} 
\cdot
C\msp{1}(n^{-1/2})^2 
\cdot 
\tP\bigl(\msp{1}j_n(w) \ge C w^2 \bigr)
\;\sim\; 
n^{-3/2}
\,.
\]
Just like with the upper bound the main contribution of $ \tE\msp{1}j_n(w) $ comes from the unlikely events, e.g., when $ \abs{X_n} \lesssim w^2 $. 
For this reason one needs again the pointwise bounds (a) and (b). 
However, unlike in \eqref{outline:upper bound first lines} the lower bound depends in a non-trivial way also on $ D_n(e_2) $ since by \eqref{def of jn} one has 
\bels{outline:importance of D_2n for the lower bound}{
\tP\bigl(\msp{1}j_n(w) \ge C_1w^2 \bigr) 
\;\sim\;
\tP\bigr(\abs{D_n(e_1;w)} \leq w^2, \abs{D_{n}(e_2;w)} \leq w\bigl)
}
Thus, to prove the lower bound one has to be able to analyze the joint behavior of the matrix elements $(D_n(e_1),D_n(e_2)) $, or equivalently, $ (X^\vartheta_n,X^0_n,\Gamma^\vartheta_n, \Gamma^0_n) $. 
These dependencies are first addressed in Subsection \ref{ssec:Joint behavior} by deriving martingale exponent representations for both $ X^\vartheta_n-X^0_n $ and $ \Gamma^0_n/\Gamma^\vartheta_n $. 
In Subsection \ref{ssec:Proof of the lower bound} these representations are used to extract (Lemma \ref{lmm:bound for K_n and L_n probabilities}) the typical joint behavior of the processes $ D(e_k) $, $ k=1,2$.   
Based on this typical behavior one is then able to construct the final bound for the right side of \eqref{outline:importance of D_2n for the lower bound}.

\section{Representation of matrix elements}
\label{sec:Representation of matrix elements}

The purpose of this section is to derive the representation \eqref{outline:decomposition D_1n and D_2n} of processes $ D(v) $, $ v \in \R^2 $, (Proposition \ref{prp:Fundamental decomposition of D_n(v)} and Corollary \ref{crl:Fundamental decompositions of D_1n and D_2n}) in terms of the Markov process $ (X_n) $ on the unit circle $ \T $. 
The first step of this derivation is to use the M\"obius transformation, associated to the average of the transfer matrix $ \tE(A_n) $, to construct $ w $-depended change-of-coordinates $ g $ which maps the evolution of the quotients $ \xi_n = D_n/D_{n-1} $ bijectively from $ \cR $ to $ \T $.
It turns out that in these new coordinates $ x = g^{-1}(\xi ) $ the noise, $ wB_n $, can be considered as a small perturbation around the zero noise evolution, which in turn is reduced to the simple shift $ x \mapsto x + \vartheta $. This is unlike in the original coordinates $ \xi \in \cR $ where the effect of noise is typically of order $ \mathcal{O}(1) $ regardless how small $ w $ is. 
The Markov process $ (X_n) $ is now defined by $ X_n := g^{-1}(D_n/D_{n-1}) $ while the representation for the matrix elements is obtained by first writing $ D_n = g(X_n) \cdots g(X_1)\cdot D_0 $ and then using the explicit knowledge of $ g $ for expanding the resulting expression w.r.t. the small disorder $ (wB_n:n\in\N)$.

The representation \eqref{outline:decomposition D_1n and D_2n} is new. Besides having the benefits already mentioned before, it also has the nice property of reducing in the zero noise case to the explicit expression $ D_{1,n} \equiv D_n = \frac{\sin \pi \vartheta (n+1)}{\pi \vartheta} $ which was already discovered by Casher and Lebowitz  (consider $1$-periodic chain in equation (3.5) in \cite{Casher-Lebowitz-71}).
The change-of-coordinates $ g $, on the other hand, is not really new as it was already discovered in a slightly different form by Matsuda and Ishii \cite{Matsuda-Ishii-1970}.
However, since our method of deriving $ g $ is different than in \cite{Matsuda-Ishii-1970} we have decided to include it here for the convenience of the reader.  

In a more general context, our representation \eqref{outline:decomposition D_1n and D_2n} is similar to some standard decomposition of products on Markov chains. 
Indeed, since $ D_n = \xi_n \cdots \xi_1 D_0 $ with $ \xi_k = D_n/D_{n-1}$, and since the transfer operator of the chain $(\xi_n) $ admits a spectral gap \cite{O'Connor-75}, a general argument \cite{Guivarch-School2002} allows us to write the decomposition $\abs{D_n} = \nE^{\gamma n + M_n}u(\xi_n)$, where $\gamma $ is a Lyapunov exponent, $ (M_n) $ is a martingale, and $ u $ is a function on $ \R $.
Although, one is not in general able to determine $ M_n $ and $ u $,  it turns out that, in the special case of random matrices, Raugi \cite{Raugi-1997} has been able to compute them explicitly, up to the knowledge of the invariant measure of the chain $(\xi_n)$. 
Still, the derivation of our formula \eqref{outline:decomposition D_1n and D_2n} is much more straightforward than the use of Raugi's formula.

\subsection{Expansion around zero noise evolution}

Let us associate a M\"obius transformation $ \mathcal{M}_A : \C \to \C $ to a $ 2 \times 2 $ to a square matrix $ A $ by setting 
\[
\mathcal{M}_A(z) \;:=\; \frac{az+b}{cz+d} \qquad \text{for}\quad A \;=\; \mat{a & b \\ c & d}
\,.
\]
The association $ A \mapsto \mathcal{M}_A $ preserves the matrix multiplication  
\bels{Mobius maps preserve matrix multiplication}{
\mathcal{M}_A \circ \mathcal{M}_B \,=\, \mathcal{M}_{AB}\,, 
\qquad (A,B \in \C^{2 \times 2}\,)
} 
so that $ (\mathcal{M}_A)^{-1} = \mathcal{M}_{A^{-1}} $ whenever either side of the equality exists.

By writing $ D_n \equiv D_n(v) $, $ v = [v_0\msp{6}v_{-1}]^\trans \in \C^2 $, and using \eqref{def of process D_n(v)} one sees that the ratios 
\bels{def of xi_n}{
\xi_n  \;:=\; \frac{D_n}{D_{n-1}}
\,,
} 
form a Markov process $ \xi \equiv (\xi_n:n\in \N_0) $ which satisfies a simple recursion relation:
\begin{subequations}
\label{system for xi_n}
\begin{align}
\label{recursion for xi_n}
\xi_n \;&=\; \mathcal{M}_{A_n}(\xi_{n-1})  
\qquad (n\in \N)
\\
\label{initial condition for xi_n}
\xi_{0} &=\; \frac{v_0}{v_{-1}}
\,.
\end{align}
\end{subequations}
Here the random matrices $ A_n $ depend on $ B_n $ through the relation \eqref{def of matrix A_k}.
Since $ \mathcal{M}_{A_n}(\pm \infty) =  2 - \pi^2 w^2 (1+B_n)$ we identify $ \pm \infty = \infty $.
By using \eqref{def of xi_n} and \eqref{system for xi_n} we get 
\bels{D_n as product of iterated fractions}{
D_{n} \;&=\; \xi_{n} \xi_{n-1} \cdots \xi_{1} D_0
\,,
}
provided no $ \xi_k \in \sett{0,\infty} $.
Recall that $ \cR $ denotes $ \R \cap \sett{\infty} $. In the following we shall consider \eqref{system for xi_n} on $ \cR $ instead on $ \C \cap \sett{\infty} $. 

\Lemma{mean diagonal coordinates}{
There exists a coordinate transformation $ g: \T \to \cR $ such that
\bels{mean evolution is shift}{
(g^{-1} \circ \mathcal{M}_{\tE(A_k)} \circ g)(x) \;=\; x \,+\, \vartheta
\msp{50}(x \in \T)
\,,
}
where $ A_k $ is the random matrix \eqref{def of matrix A_k}, and the constant shift is given by
\bels{def of average shift}{
\vartheta 
\;\equiv\; 
\vartheta(w) 
\;=\; \frac{1}{\pi}\arccos \left[1-\frac{\pi^2w^2}{2}\right] \;=\; w \,+\, \mathcal{O}(w^3) 
\,.
}
The function $  g $ and and its inverse $ g^{-1} $ are given by 
\begin{align}
\label{def of g}
g(x) \;&=\; \bigl(\mathcal{M}_{G} \circ E^{-1}\bigr)(x) 
\;=\;
\frac{ \tan \pi x}{\cos \pi \vartheta \msp{1}\tan \pi x \,+\, \sin \pi \vartheta}
\\
\label{def of inverse of g}
g^{-1}(\xi) \;&=\; \bigl(E \circ \mathcal{M}_{G^{-1}}\bigr)(\xi)
\;=\;
\frac{1}{\pi} \arctan\left[ \frac{(\sin \pi\vartheta)\, \xi}{(\cos \pi\vartheta)\, \xi \,- 1 } \right]
\,,
\end{align}
where $ E : \partial D := \sett{z \in \C : \abs{z}=1} \to \T $ is the bijection $ \nE^{\cI \phi} \mapsto \frac{\phi}{2\pi} $, and the columns of $ G $ consists of eigenvectors of $ \tE(A_l) $.    
}
\begin{Proof}
By diagonalizing, we get $ \tE(A_l) = G \Lambda G^{-1} $ where
\bels{matrices}{
\Lambda \;=\; \mat{\,\nE^{\cI \pi \vartheta} & 0 \\ 0 & \,\nE^{-\cI \pi \vartheta}}
\,, 
\quad 
G \;=\; \mat{\,1 & -1 \\ \nE^{-\cI \pi \vartheta} & -\nE^{\cI \pi \vartheta}}
\,,
\quad
G^{-1} =\; \frac{1}{2\cI \sin \pi \vartheta}\mat{\nE^{\cI \pi \vartheta} & -1\, \\ \,\nE^{-\cI \pi \vartheta} & -1}
\,,
}
and $ \vartheta $ is given in \eqref{def of average shift}. 
From \eqref{matrices} we see that $ \mathcal{M}_{G^{-1}}(\cR) = \partial D $. Since the matrix $ G $ is invertible, the property \eqref{Mobius maps preserve matrix multiplication} implies that the associated M\"obius transformation is also invertible. In particular, the restrictions $ \mathcal{M}_{G}|_{\partial D} $ and $ \mathcal{M}_{G}^{-1}|_{\cR} = \mathcal{M}_{G^{-1}}|_{\cR} $ are bijections mapping $ \partial D $ into $ \cR $ and $ \cR $ into $ \partial D $, respectively.
Using these observations we identify the coordinate transformation $ g : \T \to \cR $ and its inverse $ g^{-1} : \cR \to \T $ by regrouping as follows:
\bels{main step in identifying evolution as a shift}{
\mathcal{M}_{\tE(A_l)} \;
&=\; 
\mathcal{M}_{G} \circ \mathcal{M}_{\Lambda} \circ \mathcal{M}_{G^{-1}}
\\
&=\; 
\bigl( \mathcal{M}_{G} \circ E^{-1} \bigr) 
\circ 
\bigl( E \circ \mathcal{M}_{\Lambda} \circ E^{-1}\bigr) 
\circ 
\bigl( E \circ \mathcal{M}_{G}^{-1} \bigr)
\\
&=\; 
g \circ \lambda \circ g^{-1}
\,,
}
where $ \lambda $ equals the shift function on the right of \eqref{mean evolution is shift}. 

In order to derive \eqref{def of g} and \eqref{def of inverse of g} the easiest way is to first solve $ g^{-1} $ using $ E(z/z^\ast) = 2E(z) = \pi^{-1}\arctan\bigl[\Im(z)/\Re(z)\bigr] $: 
\[
x \;=\; g^{-1}(\xi) \;\equiv\; E\!\left( \frac{\nE^{\cI \pi \vartheta} \xi - 1}{\nE^{-\cI \pi \vartheta} \xi - 1} \right) 
\;=\;
\pi^{-1} \arctan\left[ \frac{ \xi \sin (\pi \vartheta) }{\xi \cos (\pi \vartheta)  - 1}\right] 
\,.
\]
The formula for $ g $ follows now by simply inverting the above function.
\end{Proof}

Suppose $ \xi \in \cR $ and $ \xi' = \mathcal{M}_{\tE(A_l)}(\xi) $. The important property of the new coordinates $ x $ is that even though the step $ \abs{\xi' - \xi} $ can be arbitrary large\footnote{Jumps $ \abs{\xi'-\xi} $ become arbitrary large as $ \xi $ approaches $ 0 $.} regardless of how small $ w $ is, in the new coordinates every step $ g^{-1}(\xi')-g^{-1}(\xi) = \vartheta $ is of size $ w $. The next lemma says that this property remains true even when $ \mathcal{M}_{\tE(A_l)} $ is replaced by the random evolution $ \mathcal{M}_{A_l} $.

\Lemma{functions fb and Phi}{
Let $ w > 0 $ be fixed and let $ g : \T \to \cR $ be the $ w $-dependent coordinate tranformation \eqref{def of g}. 
Then for any $ b \in \;]0,\infty[\, $ the function
\bels{definition of f_b}{
f_b \;:=\; g \circ \mathcal{M}_{A} \circ g^{-1} : \T \to \T
\qquad\text{where}\qquad 
A \;\equiv\;
A(b) \;:=\; \mat{2 - \pi^2 w^2 (1+b) & -1 \\ 1 & 0}
\,,
}
is a bijection, that can be written as
\begin{subequations}
\label{f_b and its inverse in terms of Phi}
\begin{align}
\label{f_b in terms of Phi}
f_b(x) \;&=\;  x \,+\, \vartheta \,+\, \Phi(x,b)
\\
\label{inverse of f_b in terms of Phi}
f^{-1}_b(y) \;&=\; y \,-\, \vartheta \,+\, \Phi(y-\vartheta,-b )
\,,
\end{align}
\end{subequations}
where the constant $ \vartheta = w + \mathcal{O}(w^3)$ is given in \eqref{def of average shift} and the smooth function $ \Phi : \T \times \,]0,\infty[\; \to \T $ is specified by
\begin{subequations}
\label{def of Phi}
\begin{align}
\label{Phi as arctan}
\Phi(x,b) 
\;&=\;  
\frac{1}{\pi}\arctan\left\{  \frac{ (\pi w/2) \bigl[1 - \cos (2\pi x)\bigr]\, b }{\sqrt{1-(\pi w/2)^2} \,-\, (\pi w/2) \sin (2\pi x)\,b}\right\}
\\
\label{Phi in product form}
&=\;
\sin^2 (\pi x) \Bigl[ wb \,+\, w^2b^2 (\pi/2)\sin (2\pi x) \,+\,w^3b\,R_3(w,x,b)\Bigr]
\,. 
\end{align}
\end{subequations}
The remainder term $ R_3 : [0,w_0] \times \T \times [b_-,b_+] $ is a smooth and bounded function. 
}

The lemma says that in $ x $-coordinates the system $ \xi_n = \mathcal{M}_{A_n}\msp{-2}(\xi_{n-1}) $, $ n \in \N $, and $ \xi_0 = g^{-1}(x) $ is described by the following process on a circle. The proof which is just a mechanical calculation can be found in appendix \ref{assec:Proof of fb-lemma}.

\Definition{Process X^x}{
Let $ x \in \T $. Markov process $ X^x \equiv (X^x_n:n\in\N_0) $ on $ \T $ is defined by setting
\bels{def of general process X^x}{
X^x_n \;&=\; f_{B_n}\msp{-2}(X^x_{n-1}) \qquad (n \in \N)
\\
X^x_0 \;&=\; x
\,.
} 
When the starting point $ x $ is known from the context or its specific value is not relevant we write simply $ X $ and $ X_n $ instead of $ X^x $ and $ X^x_n $, respectively. 
}
The main properties of $ f_b(x) $ are best seen by expanding it into the power series w.r.t. $ w $. Indeed, by using \eqref{def of average shift}, \eqref{f_b in terms of Phi} and \eqref{def of Phi} one gets:
\begin{subequations}
\label{expansion of fb, def of phi and psi}
\begin{align}
f_b(x) \;&=\; x \,+\, w \,+\,w \phi(x)\,b \,+\,w^2\psi(x)b^2\,+\,\mathcal{O}(w^3)
\,,   
\\
\label{def of phi}
\phi(x) \;&=\; \sin^2\msp{-1} \pi x
\,,
\\
\label{def of psi}
\psi(x) \;&=\; \pi \sin^3\msp{-1} \pi x\, \cos \pi x
\,.
\end{align}
\end{subequations}
Let us denote $ \Delta Z_k := Z_k - Z_{k-1} $ for a stochastic process $ (Z_k) $. By using the expansion \eqref{expansion of fb, def of phi and psi} together with $ \tE(B_k) = 0 $ and $ B_k \ge b_- > -1 $ the following qualitative properties of $ X $ emerge.

\Corollary{the three qualitative properties of X}{
The process $ X $ has the following three useful properties: 
\begin{itemize}
\titem{i} Uniform monotonicity: 
$ 0 \;<\; (1+b_-)w + \mathcal{O}(w^2) \;\leq\; \Delta X_k \;\leq\; (1+b_+)w + \mathcal{O}(w^2)\,; $
\titem{ii} $ \mathcal{O}(w^1) $-martingale property modulo constant shift:
$ \tE\bigl[\Delta X_k - w\big|\mathcal{F}_{k-1}\bigr] \;=\; X_{k-1} \,+\, \mathcal{O}(w^2)\,;
$ 
\titem{iii} Uniform diffusion outside any neighborhood of zero: There are constants $ \alpha(\varepsilon),\beta > 0 $ such that $\tE\bigl[(\Delta X_k-w)^2\big|X_{k-1}=x\bigr] \,\in\, [\alpha(\varepsilon) w^2, \beta w^2] $ for $ \abs{x}_\T \ge  \varepsilon $.
\end{itemize}
}
Having found good coordinates $ x = g(\xi) $ where $ \xi_n = D_n/D_{n-1} $ evolves in $ w$-sized steps in a relatively simple manner, our next step is to express the matrix elements of $ Q_n $ in terms of these new coordinates.

\Proposition{Fundamental decomposition of D_n(v)}{
Let $ v = [v_0 \msp{6} v_{-1}]^\trans \in \bar{\R}^2 $ with $ v_0 \neq 0 $.
Then there is a constant $ w_0 > 0 $ such that for $ w \in \;]0,w_0] $ the solution of \eqref{def of process D_n(v)} is
\bels{fundamental decomposition of D_n(v)}{
D_n(v) 
\;&=\; v_0 \cdot \Gamma_n^x \cdot \frac{\sin \pi X^x_n}{\sin \pi[\vartheta + \Phi(x,B_1)]}
\qquad\text{with}\quad 
x = g^{-1}(v_1/v_2)
\,,
}
almost surely.
Here the random amplitude $ \Gamma^x_n : \Omega \to \;]0,\infty[\,$ has an exponential representation 
\bels{def of Gamma^x_n}{
\Gamma^x_n \;=\; 
\exp\Biggl[ w \sum_{l=1}^n s(X_{l-1}^x)B_l \,+\, w^2  \sum_{l=1}^n r(X^x_{l-1})B_l^2 \,+\,\mathcal{O}(w^3n) \Biggr]
\,,
}
where the smooth functions $ r,s : \T \to \R $ are specified by 
\begin{subequations}
\label{s and r of Gamma}
\begin{align}
\label{function s}
s(x) \;&=\;  - \frac{\pi}{2} \sin 2\pi x
\,,
\\
\label{function r}
r(x) \;&=\; \frac{\pi^2}{4} ( \cos^2 2\pi x \,-\, \cos 2\pi x ) 
\,.
\end{align}
\end{subequations}
}

\begin{Proof}
Denote $ D_n := D_n(v) $, $ \xi_n = D_n/D_{n-1} $ and set $ x := g^{-1}(\xi_0) \equiv g^{-1}(v_0/v_{-1}) $. 
By definition \eqref{system for xi_n} the process $ (\xi_n) $ is described in $ x$-coordinates by the process $ (X^x_n) $. 
Set $ X_n := X^x_n $ and use \eqref{def of g} to write
\bels{D_l ratio in terms of X_l}{
\xi_{l} \;=\; g \circ X_{l} \;=\; \mathcal{M}_{G} \circ E^{-1}(X_{l}) \;=\; \mathcal{M}_{G}(\nE^{\cI 2\pi X_{l}}) 
\,.
}
By using \eqref{matrices} to write out the M\"obius transformation we obtain:
\[
\mathcal{M}_G(\nE^{\cI \phi}) \;=\; \frac{\nE^{\cI \phi} - 1}{\nE^{\cI (\phi-\pi\vartheta)} - \nE^{\cI \pi \vartheta}}
\;=\;
\frac{\sin \frac{\phi}{2}}{\sin \bigl(\frac{\phi}{2}-\pi \vartheta\bigr)}
\,.
\] 
By combining this with \eqref{D_l ratio in terms of X_l}, reorganizing the resulting product and then using \eqref{f_b in terms of Phi} to write $ f $ in terms of $ \Phi $ yields 
\begin{align}
\notag
\frac{D_n}{v_0} \;&=\; \frac{\xi_{n} \xi_{n-1}\cdots \xi_{1} v_0}{v_0}
\;=\; 
\prod_{l=1}^n \frac{\sin \pi X_l}{\sin \pi(X_{l}-\vartheta)}
\;=\;
\frac{\sin \pi X_n}{\sin \pi(X_1 - \vartheta)} 
\prod_{l=1}^{n-1} 
\frac{\sin \pi X_{l}}{\sin \pi (X_{l+1}-\vartheta)}
\\
\label{D_n as a product of sin-ratios}
&=\;
\frac{\sin \pi X_{n}}{\sin \pi [x + \Phi(x,B_1)]} 
\prod_{l=1}^{n-1} 
\frac{\sin \pi X_{l}}{\sin \pi [X_{l} + \Phi(X_{l},B_{l+1})]}
\,.
\end{align}
Here the possible extreme values $ \xi_k \in \sett{0,\infty} $ do not cause problems because we assumed $ \xi_0 = v_0/v_{-1} \neq 0 $ and \eqref{system for xi_n} implies
\[
\tP\bigl(\xi_k \in \sett{0,\infty} \text{ for some }k \in \N \msp{1}\big| \xi_0 \neq 0\bigr) \;=\; 0
\,.
\]
We must now show that the product of sin ratios in \eqref{D_n as a product of sin-ratios} equals the exponent $ \Gamma^x_n $. Since, the terms in the product are all similar let us consider only one such factor. 
From \eqref{Phi in product form} one sees that $ \Phi(x,b) = \mathcal{O}(w) $. This suggests expressing the denominators on the last line of \eqref{D_n as a product of sin-ratios} as power series of $ \pi \Phi(x,b) $ around zero: 
\bels{first expansion of sin-denominators}{
\sin \pi( x + \Phi(x,b))
\;&=\;
\sin \pi x\, \cos \pi\Phi(x,b) \,+\, \cos \pi x \,\sin \pi \Phi(x,b)
\\
&=\;
\sin \pi x\, \bigl\{1 -\frac{1}{2}\pi^2\Phi^2(x,b)\bigr\} \,+\, \pi \Phi(x,b) \cos \pi x \,+\, \mathcal{O}\bigl(\Phi^3(x,b)\bigr)
\,.
}
The expression \eqref{Phi in product form} also shows that $ \Phi^k(x,b)/\sin \pi x = \mathcal{O}(w^k) $ for $ k \ge 1/2 $. 
Thus using \eqref{first expansion of sin-denominators} to rewrite the denominators in \eqref{D_n as a product of sin-ratios} and then dividing the numerator and the denominator by $ \sin \pi x $ yields the expression for geometric sum of variable $ q = -\pi \Phi(x,b) \cot \pi x  + \frac{\pi^2}{2}\Phi^2(x,b) + \mathcal{O}(w^3) \,=\, \mathcal{O}(w) $. Expanding this geometric sum gives the first line of
\bea{
\frac{\sin \pi x}{\sin \pi(x +\Phi(x,b))}\;
&=\;
1 \,-\,\pi \Phi(x,b)\cot \pi x  + \frac{\pi^2}{2}\Phi^2(x,b) \,+\,\pi^2 \Phi^2(x,b) \cot^2 \!\pi x
\,+\,\mathcal{O}(w^3)
\\
&=\;
1 \,-\, w\frac{\pi}{2} \sin 2\pi x\;b \,+\,w^2 \frac{\pi^2}{8}(1-\cos 2\pi x)^2\,b^2 + \mathcal{O}(w^3)
\,,
}
while the last line follows from \eqref{Phi in product form} and trigonometric double angle formulae. By using $  1+z = \exp \circ \ln \msp{1} (1+z) = \exp\bigl[ z - \frac{1}{2}z^2 + \mathcal{O}(z^3) \bigr] $, with $ \abs{z} \leq Cw_0 $, for the last expression we get
\bea{
\frac{\sin \pi x}{\sin \pi (x +\Phi(x,b))}\;
&=\;
\exp\Bigl[ -w (\pi/2) \sin 2\pi x\; b \,+\,w^2 (\pi/2)^2\bigl( \cos^2 2\pi x - \cos 2\pi x \bigr) b^2 + \mathcal{O}(w^3) \Bigr]
\,.
}
Identifying functions $ s $ and $ r $ on the right side and then applying this bound term by term for the product in \eqref{D_n as a product of sin-ratios} yields the expression on the right side of \eqref{def of Gamma^x_n}.
\end{Proof}

It is worth remarking that the proposition does not apply directly for $ v \in \C^2 $ since it relies on Lemmas \ref{lmm:mean diagonal coordinates} and \ref{lmm:functions fb and Phi} which apply only when $ (\xi_n) $ takes values on $ \R $. Of course, by the linearity of the system \eqref{def of process D_n(v)} one still has $ D_n(v_R + \cI v_I) = D_n(v_R) + \cI D_n(v_I) $ for any $ v_R,v_I \in \R^2 $. 
The next corollary shows that the generic choice $ D_n(v) $ with $ v = e_k $, $ k =1,2$, is often a convenient choice as $ D(e_2)$ can be treated as a perturbation of $ D(e_1) $. 

\Corollary{Fundamental decompositions of D_1n and D_2n}{
There is a constant $ w_0 > 0 $ such that for $ w \in \;]0,w_0] $:
\begin{subequations}
\label{fundamental decompositions of D_1n and D_2n}
\begin{align}
\label{fundamental decomposition of D_1n}
D_n(e_1) 
\;&=\; \Gamma_n^\vartheta \cdot \frac{\sin \pi X^{\vartheta}_n}{\sin \pi[\vartheta + \Phi(\vartheta,B_1)]}
\;\sim\;
w^{-1} \Gamma_n^\vartheta\cdot \sin \pi X^\vartheta_n
\,,
\\
\label{fundamental decomposition of D_2n}
D_n(e_2) 
\;&=\; \Gamma_n^0 \cdot \frac{\sin \pi X^0_n}{\sin \pi[\vartheta + \Phi(\vartheta,B_2)]}
\;\sim\;
w^{-1} \Gamma_n^0 \cdot \sin \pi X^0_n
\,.
\end{align}
\end{subequations}
}

\begin{Proof}
By \eqref{def of inverse of g} we get $ g^{-1}(\xi_0) = g^{-1}(1/0) = \vartheta $ and thus \eqref{fundamental decomposition of D_1n} follows directly from Proposition \ref{prp:Fundamental decomposition of D_n(v)}. 
In order to prove \eqref{fundamental decomposition of D_2n} one can not directly apply the proposition since the first component of $ e_2 $ is zero. However, from \eqref{def of process D_n(v)} one sees that $ [D_1(e_2) \msp{8}D_0(e_2)]^\trans = [-1 \msp{8}0]^\trans = -e_1 $ and $ D_n(-v) = -D_n(v) $. 
Thus, by defining $ \theta : \Omega \to \Omega $ by $ \theta\omega = (b_2,b_3,\dots) $ for $ \omega = (b_1,b_2,\dots) $ and denoting the associated pullback $ \theta_\ast $ on random variables $ Z $ by $ \theta_\ast Z(\omega) = Z(\theta \omega) $, one can write
\bels{representation for D_2n - step 1}{
D_n(e_2) 
\,=\; - \theta_\ast D_{n-1} 
\;=\; 
\theta_\ast\Gamma^\vartheta_{n-1} \cdot \frac{\sin \pi \msp{1}\theta_\ast\msp{-1} X^\vartheta_{n-1}}{\sin \pi[\vartheta + \Phi(\vartheta,\theta_\ast\msp{-1} B_1)]} 
\,,
}
where by the definition:
\bels{finding Gamma^0_n - step 1}{
\theta_\ast\Gamma^\vartheta_{n-1}
\;=\;
\exp\Biggl[ w \sum_{l=1}^{n-1} s(\theta_\ast X^\vartheta_{l-1})\,\theta_\ast \msp{-1}B_l \,+\, w^2  \sum_{l=1}^{n-1} r(\theta_\ast X^\vartheta_{l-1})(\theta_\ast\msp{-1} B_l)^2 \,+\,\mathcal{O}(w^3n) \Biggr]
\,.
}
Now, since $ \Phi(0,b) = 0 $ it follows that $ X^0_1 = f_{B_1}(0) = \vartheta  + \Phi(0,B_1) = \vartheta = \theta_\ast X^\vartheta_0 $ regardless of the value of $ B_1 $. But $ (X^0_n:n\in\N) $ and $ (\theta_\ast X^\vartheta_{n-1}:n\in\N) $ also satisfy the same recursion relations for $ n \ge 2 $ and therefore $ \theta_\ast X^\vartheta_n = X^0_{n+1} $, $ n \in \N_0 $. Also, by definition $ \theta_\ast B_l(\omega) = b_{l+1} = B_{l+1}(\omega) $.
Thus we may replace $ \theta_\ast X^\vartheta_{l-1} $ with $ X^0_l $ and write $ \theta_\ast B_l = B_{l+1} $ in \eqref{representation for D_2n - step 1} and \eqref{finding Gamma^0_n - step 1}. Moreover, if we also reindex the sums in \eqref{finding Gamma^0_n - step 1} we obtain an exponential representation for $ \theta_\ast \Gamma_{n-1} $ that is up to a missing first terms $ w\msp{1}s(X^0_0)B_1 $ and $ w^2r(X^0_0)B_1^2 $ equal to $ \Gamma^0_n $. However, these missing terms are both zero due to the "coincidence" $ s(0) = r(0) = 0 $, and thus we get $ \theta_\ast\Gamma_{n-1} = \Gamma^0_n $. This proves \eqref{fundamental decomposition of D_2n}.
\end{Proof}

\subsection{Joint behavior}
\label{ssec:Joint behavior}

In order to prove $ n^{-3/2} \lesssim J_n $ we analyze the current density $ j_n $ defined in \eqref{def of jn}. This leads us to consider the properties of the  quadruple $ (X^\vartheta_n,X^0_n,\Gamma^\vartheta_n,\Gamma^0_n) $. 
Since $ X^\vartheta_0-X^0_0 = \vartheta \sim w $ one can consider $ X^0_n $ and $ \Gamma^0_n $ as perturbations around $ X^\vartheta_n $ and $ \Gamma^\vartheta_n $, respectively. Based on this simple idea one proves the following. 

\Lemma{Representations for the lower bound}{
Let us treat $ X^x $, $ x \in \R $ as real valued processes. Then for all $ n \in \N $ and $ w \in \;]0,w_0] $:
\begin{align}
\label{R-distance between X^vartheta_n and X^0_n}
X^\vartheta_n -\, X^0_n \;=&\; w\, \nE^{M_n+\,L_n+\,\mathcal{O}(w^2n)}
\\
\label{Gamma_2n from Gamma_1n}
\Gamma^0_n/\Gamma^\vartheta_n \;=&\; \nE^{K_n +\, \mathcal{O}(w+w^2n)}
\,,
\end{align}
where $ (M_n), (L_n), (K_n) $ are $ \R$-valued $ \mathbb{F} $-martingales such that $ M_0 = L_0 = K_0 = 0 $ and $ n \in \N $:
\begin{align}
\label{def of dM_n}
\Delta M_n \;&=\; w\msp{1} \phi'(X^\vartheta_{n-1}) B_n
\\
\label{def of dL_n}
\Delta L_n \;&=\; 
w^2 \nE^{M_{n-1}+\,L_{n-1}+\,\mathcal{O}(w^2n)} H_{n-1} B_n
\\
\label{def of dK_n}
\Delta K_n \;&=\; 
w^2 \nE^{M_{n-1}+\,L_{n-1}+\,\mathcal{O}(w^2n)} U_{n-1} B_n
\,.
\end{align}
The processes $ (H_n) $ and $ (U_n) $ are $ \mathbb{F} $-adapted and bounded such that:
\bels{increments bounded by constant}{
\sup \;\setb{\abs{H_n},\,\abs{U_n},\,w^{-1}\msp{-1}\abs{\Delta L_n},\, w^{-1}\msp{-1}\abs{\Delta K_n}\,:\,n \in \N} \;&\leq\; C\,. 
}
}

\begin{Proof}
From \eqref{Phi in product form} and \eqref{def of phi} one sees that $ \Phi(x,b) = w \phi(x)b + w^2 R_2(x,b) $ where $ R_2 $ is a smooth and bounded function. Using \eqref{f_b in terms of Phi} we get 
\bels{f_b difference 1}{
f_b(x)\,-\,f_b(x-z) \;
&=\;z \,+\,\Phi(x,b)-\Phi(x-z,b)
\\
&=\;\biggl\{ 1 \,+\,w \frac{\phi(x)-\phi(x-z)}{z}\,b \,+\,w^2 \frac{R_2(x,b)-R_2(x-z,b)}{z}\biggr\}\, z 
\,,
}
for any $ z \in \R  $. By the mean value theorem there are function $ \zeta_1(x,z), \zeta_2(x,z,b) \in [x-z,x] $ such that for any $ x \in \R $, $ z \ge 0 $ and $ b \in [b_-,b_+]$ we have
\bels{f_b difference 2}{
f_b(x)\,-\,f_b(x-z) 
\;&=\;
\biggl\{1 \,+\, w \phi'(x)b \,-\,w z\,\frac{1}{2}\phi''(\zeta_1(x,z))\msp{1} b\,+\,
w^2 \partial_x R_2(\zeta_2(x,z,b),b) 
\biggr\}\, z
\\
&=\; \exp \biggl[w \phi'(x)b \,-\, wz\,\frac{1}{2}\phi'' \!\circ \zeta_1(x,z)\msp{1} b \,+\, \mathcal{O}(w^2) \biggr]\, z
\,.
}
Now, set 
\bels{Theta_n and H_n explicitly}{
\Theta_n \;:=\; 
(X^{\vartheta}_n - X^0_n) / w 
\qquad\text{and}\qquad  
H_n \;:=\; -\frac{1}{2} \phi'' \!\circ \zeta_1(X^\vartheta_n,w\msp{1}\Theta_n)
\,,
}
Then \eqref{f_b difference 2} and \eqref{def of general process X^x} yield
\begin{align}
\notag
\Theta_n \;
&=\; \frac{1}{w}\bigl\{f_{B_n}\!(X^\vartheta_{n-1}) \,-\, f_{B_n}\!(X^\vartheta_{n-1}\!-w\Theta_{n-1})\bigr\}
\\
\notag
&=\; \exp \Bigl[ w \phi'(X^\vartheta_{n-1}) B_n \,-\, w^2\Theta_{n-1}\frac{1}{2}\phi''\! \circ \zeta_1(X^\vartheta_{n-1},w\Theta_{n-1})\msp{1} B_n \,+\,\mathcal{O}(w^2)\Bigr] \cdot \Theta_{n-1} 
\\
\label{exponent representation 1 for Theta_n}
&=\; \exp \Biggl[ w \sum_{j=1}^n \phi'(X^\vartheta_{j-1}) B_j \,+\, w^2 \sum_{j=1}^n \Theta_{j-1}\,H_{j-1}B_j \,+\, \mathcal{O}(w^2n) \Biggr] \cdot \Theta_0
\,.
\end{align}
By using \eqref{def of dM_n} and \eqref{def of dL_n} we identify the two sums inside the exponent in \eqref{exponent representation 1 for Theta_n} as $ M_n $ and $ L_n $, respectively. Together with $ \Theta_0 = (X^\vartheta_0-X^0_0)/w = \vartheta/w = 1 + \mathcal{O}(w^2) $ this gives $ \Theta_n = \nE^{M_n + L_n + \mathcal{O}(w^2n)} $ and by the definition \eqref{Theta_n and H_n explicitly} this equals \eqref{R-distance between X^vartheta_n and X^0_n}. 
Moreover, $ w^{-1} \Delta L_{n+1} = w \Theta_nH_n B_{n+1} $, where using \eqref{f_b difference 2}, \eqref{Theta_n and H_n explicitly} and the definition of $ \zeta_1 $ we get  
\[
w \msp{1} \Theta_n H_n 
\;=\; 
\frac{\phi(X^\vartheta_n)-\phi(X^0_n)}{X^\vartheta_n-X^0_n} \,-\, \phi'(X^\vartheta_n) 
\;=:\; 
\phi'(\zeta_0) \,-\, \phi'(X^\vartheta_n)
\,,
\]
for some $ \zeta_0 \in [X^0_n,X^\vartheta_n] $, and therefore $ w^{-1} \abs{\Delta L_{n+1}} \leq 2 \norm{\phi'}_\infty \!\cdot \max\sett{-b_-,b_+} =: C $.

In order to prove \eqref{Gamma_2n from Gamma_1n} we use again the mean value theorem to write
\bels{s at X^0_n as an expansion around X^vartheta_n}{
s(X^0_n) 
\;=\; 
s(X^\vartheta_n-w\msp{1}\Theta_n) 
\;=\; 
s(X^\vartheta_n) \,-\, w\msp{1}\Theta_n\cdot s' \circ \zeta_3(X^\vartheta_n,w\msp{1}\Theta_n)
\,,
}
where $ X^\vartheta_n - w\Theta_n \leq \zeta_3(X^\vartheta_n,w\Theta_n) \leq  X^\vartheta_n $. Using this in \eqref{def of Gamma^x_n} yields 
\bea{
\Gamma^0_n \;
&=\; \exp\left[w\sum_{l=1}^n s(X^0_{l-1})B_l \,+\, \mathcal{O}(w^2n)\right] 
\\
&=\; \exp\left[w\sum_{l=1}^n s(X^\vartheta_{l-1})B_l \,-\,w^2\sum_{l=1}^n \Theta_{l-1} \msp{-2}\cdot s' \circ \zeta_3(X^\vartheta_{l-1},w\Theta_{l-1})\msp{1}B_l \,+\, \mathcal{O}(w^2n)\right] 
\\
&=:\; 
\Gamma^\vartheta_n\, \nE^{K_n \,+\, \mathcal{O}(w+w^2n)} 
\,.
}
Above, we have identified $ U_n = -s' \circ \zeta_3(X^\vartheta_n,w\msp{1}\Theta_n) $ in \eqref{def of dK_n}. Finally, by equation \eqref{s at X^0_n as an expansion around X^vartheta_n}
$ w^{-1} \Delta K_{n+1} = w\Theta_n U_n B_{n+1} = [s(X^0_n)-s(X^\vartheta_n)]\, B_{n+1} $. Since $ s $ is a bounded function \eqref{function s} this implies $ w^{-1}\abs{\Delta K_n} \leq C $.
\end{Proof}

\section{Expectation of $ 1/\Gamma_n $}
\label{sec:Expectation of 1/Gamma_n}

In this section we prove the following result. 
\Proposition{expectation of 1/Gamma_n decays exponentially}{
For sufficiently small $ w_{0} \sim 1 $ there exists $ \alpha \equiv \alpha(w_0) > 0 $ such that for $ n \in \N $,
\bels{expectation of 1/Gamma_n decays exponentially}{
\sup_{x\in\T} \tE\bigl(1/\Gamma^x_n\bigr) 
\;\lesssim\; 
\nE^{-\alpha  w^2 n}
\,,\qquad w \in \;]0,w_0]
\,.
}
}

The content of this result is best understood by using \eqref{def of Gamma^x_n} to write $
1/\Gamma_n  $ as exponent $ \nE^{-R_n w^2n \,+\, wn^{1/2}S_n \,+\, \mathcal{O}(w^3n)} $, where the normalized random variables
\[
S_n \;=\; \frac{-1}{n^{1/2}}\sum_{k=1}^n s(X_{k-1})B_k 
\qquad\text{and}\qquad
R_n \;=\; \frac{1}{n}\sum_{k=1}^n r(X_{k-1})B_k^2
\,,
\]
are in average of order $ 1 $. Our proof of Proposition \ref{prp:expectation of 1/Gamma_n decays exponentially} consists of two steps which both rely on the fact that during any consecutive sequence of $ \floor{1/w} $ steps the random set $ \{ X_j(w) : j=k,\dots,k+\floor{1/w} \} $, $ k \in \N $, typically samples $ \T $ evenly.
First, Lemma \ref{lmm:noisy Lp-ergodic over one round} is used to shows that $ R_n \equiv R_n(w) $ can be replaced by the constant $ \gamma(w)/w^2 $  without introducing too large errors in $ \tE(1/\Gamma_n) $ provided $ wn \to \infty $. Here
\bels{Lyapunov exponent explicitly solved}{
\gamma(w) 
\;=\; 
\left\{ \tE(B^2_1) \cdot \!\int_\T r(x)\dif x \right\} w^2 +\, \mathcal{O}(w^3) 
\;=\; \frac{\pi^2\tE(B^2_1)}{8}w^2 + \mathcal{O}(w^3)
\,,
}
is the Lyapunov exponent associated to the norm of $ Q_n $ in \eqref{solution of Q_n in terms of D_n(v)}. 
Secondly, the uniform monotonicity (property (i) of Corollary \ref{crl:the three qualitative properties of X}) of the process $ X $ is used to bound the conditional variance (see \eqref{def of V_n}) of the martingale $ n^{1/2}S_n $ so that Freedman's powerful exponential martingale bound, i.e., Lemma \ref{lmm:Freedman's and Azuma's martingale exponent bounds}, can be applied to obtain a bound $ \tE \msp{1}\nE^{wn^{1/2}S_n} \leq \nE^{\beta w^2n} $, where $ \gamma(w)/w^2 -\beta =: \alpha \sim 1 $.  

The following lemma provides two powerful exponential martingale bounds due to Freedman \cite{Freedman-75} and Azuma \cite{Azuma-1967}. 

\Lemma{Freedman's and Azuma's martingale exponent bounds}{
Let $ (M_i) $ be a $ (\mathcal{F}_i) $-martingale, and define a process $ (V_n) $ by setting 
$ V_0 = 0 $ and 
\bels{def of V_n}{
V_n \;:=\; \sum_{i=1}^n\tE\bigl[(M_i-M_{i-1})^2\big| \mathcal{F}_{i-1}\bigr]
\,,\qquad n\in\N
\,.
}
Suppose there exists a constant $ m $ and a sequence $ (v_n) \subset [0,\infty[\, $ such that $\abs{M_n -M_{n-1}} \leq m $ and $ V_n \leq v_n $ for all $ n \in \N $.
Then for any $ t \in \R $ and $ n \in \N $:
\begin{align}
\label{Freedman's and Azuma's exponent bounds}
\tE\, \nE^{t M_n} 
\;\leq\; 
\begin{cases}
\nE^{\msp{1}\kappa_m(t)\msp{1}v_n}
\,,\quad 
&\text{''Freedman's bound'';}
\\
\nE^{\frac{t^2}{2}m^2n}  
\,,
&\text{''Azuma's bound'';}
\end{cases}
\end{align}
where 
\bels{def of kappa_m}{
\kappa_m(t) \;=\; \frac{\nE^{mt} - 1 - mt}{m^2} \;\leq\; \frac{t^2}{2} \,+\,\frac{m}{6}\nE^{m\abs{t}}\abs{t}^3
\,.
}
}

For the convenience of readers the proofs of these bounds are included in Appendix \ref{assec:Proofs of Freedman's and Azuma's bounds}. The next inequality \eqref{Azuma's inequality} is often referred as Azuma's inequality. 

\Corollary{Azuma's inequality}{
Suppose $ (M_k) $ satisfies the hypothesis of Lemma \ref{lmm:Freedman's and Azuma's martingale exponent bounds}. Then for any $ n \in \N $ and $ r > 0 $:
\bels{Azuma's inequality}{
\tP( \abs{M_n} \ge r ) \;
&\leq\; 
2\, \nE^{-\frac{r^2}{2m^2n}}
\,.
} 
}
\begin{Proof}
The proof follows by using Markov's inequality: $ \tP(\abs{M_n} \ge r) = \tP( M_n \ge r) + \tP( -M_n \ge r) \leq \nE^{-sr} \tE\, \nE^{sM_n} + \nE^{-sr} \tE \,\nE^{-sM_n} $, and then use Azuma's bound \eqref{Freedman's and Azuma's exponent bounds} with $ t = r/(m^2n) $.
\end{Proof}

\Lemma{noisy Lp-ergodic over one round}{
Suppose $ u $ is a Lipshitz-function on $ \T $, i.e., there is a constant $ L_u > 0 $ such that for all $ x,y \in \T $: $ \abs{u(x)-u(y)} \leq L_u \abs{x-y}_\T $. Then:
\bels{Lp-ergodic average}{
\sup_{x\in\T}\tE \Biggl\{ \biggl| w \sum_{j=0}^{\floor{1/w}} u(X^x_j) \,-\, \int_\T u(y) \dif y \biggr|^p \Biggr\} \;\leq\; C_p L_u^p w^{p/2}
\,,
}
where $ C_p $ does not depend on $ u $. 
}
\begin{Proof}
Fix $ x $ and set $ X := X^x $ and $ I_j := [x+w\,(j-1),x+w\,j[\, $. Define for each $ j $ some $ \tilde{x}_j \in I_j $ by requiring $ \int_{I_j} u(x) \dif x = w\, u(\tilde{x}_j) $, and set $ \bar{x}_j := \tE(X_j) $. The properties \eqref{expansion of fb, def of phi and psi} of the chain $ X $ imply $ \abs{\bar{x}_j-\tilde{x}_j} \leq w $ for all $ j \leq \floor{1/w}$. 
By writing the integral on the left side of \eqref{Lp-ergodic average} as a sum over $ u(\tilde{x}_j) $ and then applying the Lipshitz-property of $ u $ one gets 
\bels{Lp-erg A}{
\tE \Biggl\{ \biggl| w \sum_{j=0}^{\floor{1/w}} \bigl[u(X_j) -u(\tilde{x}_j)\bigr] \biggr|^p \Biggr\}
\;\leq\;
L_u^p\, w^p \sum_{j_1,\dots,j_p} \tE \Biggl\{ \prod_{l=1}^p \abs{X_{j_l}-\tilde{x}_{j_l}}\Biggr\}
\,.
}
Now, $ X_j = x + wj + w^{1/2}M_j + \mathcal{O}(w) $ with $ M_j = w^{1/2}\sum_{i=1}^j \phi(X_{i-1})B_i $ uniformly for any $ 0 \leq j \leq \floor{1/w} $.
This means $ X_j - \tilde{x}_j = w^{1/2}(M_j + \mathcal{O}(w^{1/2})) $. By applying the generalized H\"older's inequality 
one has,
\bels{Lp-erg B}{
\tE \Biggl\{ \prod_{l=1}^p \abs{X_{j_l}-\tilde{x}_{j_l}}\Biggr\}\;
&=\;
w^{p/2} 
\tE \Biggl\{ \prod_{l=1}^p \absb{M_{j_l} + \mathcal{O}(w^{1/2})}\Biggr\}
\\
&\leq\;
w^{p/2} 
\left(\prod_{l=1}^p \tE \Bigl\{ \absb{M_{j_l} + \mathcal{O}(w^{1/2})}^p \Bigr\}\right)^{1/p}
.
}
The last expectations of \eqref{Lp-erg B} can be bounded with Azuma's inequality \eqref{Azuma's inequality}. Indeed, $ \abs{M_j-M_{j-1}} \leq w^{1/2}\max(-b_-,b_+) \norm{\phi}_\infty \equiv Cw^{1/2} $ for each $ j $. This implies $ \tP\bigl(\abs{M_j} \in [k,k+1[\,\bigr) \leq 2 \tP(\abs{M_j}\ge k) \leq 2 \nE^{-k^2/(2C^2w\floor{1/w})} = \nE^{-k^2/C'} $ which, in turn, yields
\[
\tE \Bigl\{ \absb{M_j + \mathcal{O}(w^{1/2})}^p \Bigr\} 
\;\leq\;
\sum_{k=0}^\infty (k+1+\mathcal{O}(w^{1/2}))^p \tP\bigl(\abs{M_j} \in [k,k+1[\,\bigr)
\;\leq\; 2\sum_{k = 0}^\infty k^p \nE^{-k^2/C'} \;=: C_p
\,, 
\]
Since this bound holds uniformly for all $ j = 0,1, \dots, \floor{1/w} $ we may apply it term by term in \eqref{Lp-erg B}. Using the resulting bound again term by term in \eqref{Lp-erg A} yields the bound \eqref{Lp-ergodic average}.
\end{Proof}

\begin{Proof}[Proof of Proposition \ref{prp:expectation of 1/Gamma_n decays exponentially}]
Since $ \Gamma_n(w) \ge C $ for $ wn \sim 1 $ it is enough to show $ \tE(1/\Gamma^x_n) \leq C\,\nE^{-\alpha w^2 n} $ for $ n = \floor{1/w} m $, $ m \in \N $. Since $ \Delta X_n \ge Cw $ we may for the same reason fix some arbitrary starting point $ x \in \T $ and denote $ X^x_n $ and $ \Gamma^x_n $ by $ X_n $ and $ \Gamma_n $, respectively.  
We begin the proof by decomposing the second sum in the exponent of \eqref{def of Gamma^x_n} into the double sum 
\bels{decomposition of r sum}{
w^2 \sum_{i=1}^n r(X_{i-1})B_i^2 \;
&=\;w\sum_{k=1}^m \;w \msp{-16}\sum_{i= i_{k-1}+1}^{i_{k}} \msp{-10}r(X_{i-1})B_i^2
\;=\; 
w \sum_{k=1}^m \gamma(X_{i_{k-1}}) \,+\,w\sum_{k=1}^m Z_k
\,,
}
where $ i_k = \floor{1/w}k + 1 $, $ k = 1,2,\dots,m $ is roughly the time the averaged process $ \bar{x}_j := \tE_x(X_j) = x + wj + \mathcal{O}(w^3j) $ has passed its starting point k\textsuperscript{th} time.
In the rightmost expression of \eqref{decomposition of r sum} we have further divided the inner sums into the conditional expectations and the fluctuation parts:
\begin{subequations}
\begin{align}
\label{def of Z_k}
Z_k \;:=&\; w\msp{-10}\sum_{i= i_{k-1}+1}^{i_{k}} \!r(X_{i-1})B_i^2 \,-\, \gamma(X_{i_{k-1}}) 
\\
\label{def of gamma}
\gamma(y)
\,:=&\; 
\tE\Biggl\{ w \sum_{i=1}^{\floor{1/w}} r(X_{i-1}^y) B_i^2 \Biggr\} 
\,.
\end{align}
\end{subequations}
The motivation behind the decomposition \eqref{decomposition of r sum} is twofold. First, Lemma \ref{lmm:noisy Lp-ergodic over one round} tells us that the function $ \gamma $ is almost constant for small $ w $, and especially
\bels{bound of gamma_-}{
\gamma(y) \;
&=\;
\tE(B^2)\,
\tE\Biggl\{ w\!\sum_{i=1}^{\floor{1/w}} r(X^y_{i-1})\Biggr\}
\;\ge\; 
\tE(B^2)\,\int_\T r(z) \dif z \,-\, \beta_0 w^{1/2} 
\;=:\; \tilde{\gamma}_-
\,,
} 
where $ \beta_0 > 0 $ is a finite constant that does not depend on $ y $.
Here the first equality follows from $ \tE\bigl(r(X_{i-1})B_j\bigr) = \tE(B^2)\, \tE( r^2(X_i) )$, while the last expression comes from Lemma \ref{lmm:noisy Lp-ergodic over one round} with $ p = 1 $ and $ L_u := \norm{r'}_\infty $. 
Using \eqref{bound of gamma_-} to bound each term $ \gamma(X_{i_{k-1}}) $ in \eqref{decomposition of r sum} yields the bound:
\bels{1/Gamma_n two martingales left}{
\tE\bigl(1/\Gamma_n\bigr) \;
&\leq\; 
\nE^{-\gamma_- w^2n} \tE \exp\Biggl[ -w \sum_{i=1}^n s(X_{i-1})B_i \,-\, w\sum_{k=1}^m Z_k \Biggr]
\,,
\quad\text{with}\quad
\gamma := \tilde{\gamma} + \mathcal{O}(w)
\,,
}
where the $ \mathcal{O}(w^3n) $-term inside the exponent \eqref{def of Gamma^x_n} of $ \Gamma_n $ has been also absorbed into the constant $ \gamma_- $. 

The second property of the decomposition \eqref{decomposition of r sum} is that $ (Z_k:k\in\N) $ constitutes a sequence of bounded martingale increments in the \emph{sparse filtration} $ \mathbb{F}' = (\mathcal{F}'_k) $, $ \mathcal{F}'_k := \mathcal{F}_{i_k} \equiv \sigma(B_1,B_2,\dots,B_{i_k}) $: the boundedness of $ Z_k $ is obvious as it is an average of $ \floor{1/w} $ uniformly bounded increments, while the martingale property holds, since $ X $ is Markov: 
\[ 
\tE \Biggl( w\msp{-18}\sum_{\msp{12}i=i_{k-1}+1}^{i_k} \msp{-10}r(X_{i-1}) B_i^2 \Bigg|\mathcal{F}'_{k-1} \Biggr)(\omega) 
\;=\; 
\tE\Biggl\{w\sum_{i=1}^{\floor{1/w}} r\Bigl(X^{X_{i_{k-1}}\!(\omega)}_{i-1}\Bigr) B_i^2\, \Biggr\} 
\;\equiv\; \gamma(X_{i_{k-1}}(\omega)) 
\,,
\] 
for a.e. $ \omega \in \Omega $.
We want to consider both sums in the right side of \eqref{1/Gamma_n two martingales left} as martingales. Since this is not possible under the same expectation we apply H\"older's inequality to divide the expectation into the product of separate expectations
\bels{expectation of Gamma by Holder}{
\tE (1/\Gamma_n) \;
&\leq\;
\nE^{-\gamma_{-} w^2n} 
\Biggl\{\tE \exp \biggl[ -pw \sum_{i=1}^n s(X_{i-1})B_i\biggr]\Biggr\}^{1/p} 
\Biggl\{\tE \exp \biggl[ -p'w\sum_{k=1}^m Z_k\biggr] \Biggr\}^{1/p'} 
\,,
}
where $ p,p' \ge 1 $ and $ 1/p + 1/p' = 1 $. 
We can now bound both of these expectations with the help of Lemma \ref{lmm:Freedman's and Azuma's martingale exponent bounds}. Azuma's exponential bound \eqref{Freedman's and Azuma's exponent bounds} is sufficient for the second factor: if $ \abs{Z_k} \leq C_Z $, then 
\bels{final bound for Z-exponent}{
\Biggl\{\tE \exp \biggl[ -p'w\sum_{k=1}^m Z_k\biggr] \Biggr\}^{1/p'}
\msp{-6}\leq\; 
\Biggl\{\exp\biggl[\frac{(-p'w)^2}{2}C_Z^2\floor{nw}\biggr]\Biggr\}^{1/p'}
\msp{-6}\leq\; 
\nE^{\beta_2 p' w^3n}
\,,
}
for some constant $ \beta_2 $.

In order to handle the first expectation of \eqref{expectation of Gamma by Holder} we note that the martingale $ (M_j) $, defined by $ \Delta M_j := s(X_{j-1})B_j $, $ j \in \N $ and $ M_0 = 0 $, has bounded increments. Moreover, since $ \tE[(\Delta M_i)^2|\mathcal{F}_{i-1}] = \tE(B^2) \cdot s^2(X_{i-1}) $, we see that for sufficiently small $ \varepsilon > 0 $: 
\bea{
V_n 
\;:=\; 
\sum_{i=1}^n\tE\bigl[(M_i-M_{i-1})^2\big| \mathcal{F}_{i-1}\bigr]
\;=\;
\tE(B^2) \sum_{i=1}^n s^2(X_{i-1}) \;\leq\; (1-\varepsilon)\tE(B^2)\norm{s}_\infty^2 n 
\,.
}
In order to get the last bound above, one uses the property (i) of Corollary \ref{crl:the three qualitative properties of X}, the continuity of $ s $ and $ s(0)= 0 $, to conclude that there must exist $ \varepsilon > 0 $ such that  
\[
\absb{\bigl\{0 \leq i \leq n-1: \abs{s^2(X_i)} \leq \norm{s}_\infty^2/2 \bigr\}} \;\ge\; 2\varepsilon \msp{1}n
\,.
\]
This, by definition, implies the bound of $ V_n $ above.
Applying Freedman's bound of Lemma \ref{lmm:Freedman's and Azuma's martingale exponent bounds} with $ v_n := \tE(B^2)(1-\varepsilon)\norm{s}_\infty^2 n $ and $ \abs{M_i-M_{i-1}} \leq C_M =: m $ yields
\bels{final bound for s-exponent}{
\Biggl\{\tE \exp \biggl[ (-wp) \sum_{i=1}^n s(X_{i-1})B_i\biggr]\Biggr\}^{1/p}
\;&\leq\;
\Biggl\{ \exp\biggl[\kappa_{C_M}(-wp)\tE(B^2)(1-\varepsilon)\norm{s}_\infty^2 n \biggr]\Biggr\}^{1/p}
\\
&\leq\;
\nE^{\frac{1}{2}pw^2(1-\varepsilon)\tE(B^2)\norm{s}_\infty^2n \,+\, 
\beta_1 \msp{1}p^2w^3n
}
\,,
}
where $ \beta_1 > (1/6)(1-\varepsilon)\tE(B^2)C_M \nE^{C_Mpw} \sim 1 $.

Plugging \eqref{final bound for s-exponent} and \eqref{final bound for Z-exponent} along with the estimate \eqref{bound of gamma_-} for $ \gamma_- $ into \eqref{expectation of Gamma by Holder} results into the total bound
\bels{collection of terms for 1/Gamma_n expectation}{
\tE(1/\Gamma_n) 
\;\leq\;
\nE^{-\tE(B^2)\bigl\{\int_\T r(y) \dif y \,-\, p\msp{1}(1-\varepsilon)\frac{\norm{s}_\infty^2}{2} \bigr\}w^2n \,+\, \beta_0 w^{5/2}n \,+\,\beta_1 p^2w^3n\,+\, \beta_2p'w^3n+ Cw^3n}
\,.
}
Here the term inside curly brackets would disappear if $ p=1, \varepsilon = 0 $ because 
$ \int_\T r(y) \dif y = \norm{s}_\infty^2/2 = \pi^2/8 $. However, since $ \varepsilon > 0 $ we can take $ p > 1 $ such that it remains positive. 
However, by taking $ w_0 $ sufficiently small the last three terms, regardless of the size of $ p' $ or $ \beta_1,\beta_2,\beta_3,C $, can be made arbitrary small compared to the first part. 
%
\end{Proof}

\section{Potential theory}
\label{sec:Potential theory}

This section is devoted to the statement and the proof of  Proposition \ref{prp:Potential theory} below.
The derivation of the inequalities \eqref{upper bound} and \eqref{lower bound} constitutes a relatively classical problem in potential theory for Markov chains. 
However, it does not seem possible to apply classical results (see e.g. \cite{Coulhon-1990} and \cite{Coulhon-1993}), since the chain $X$ is neither reversible, nor uniformly diffusive.
In particular, little appears to be known on lower bounds of the type \eqref{lower bound} for non-reversible Markov chains. 
Results for Markov chains on a lattice \cite{Mustapha-2006}, or for differential equations in non-divergence form \cite{Escaurazia-2000}, do not adapt straightforwardly (and maybe not at all) to our case.
Instead, since we consider only the case $w\to 0$, 
it has been possible to treat the left hand side of \eqref{upper bound} and \eqref{lower bound}
as a perturbation of quantities that can be computed explicitly.
We are then able to handle both of these bounds with a single method. 

\Proposition{Potential theory}{
Let $\kappa>0$, and let $ h \in \Cont^1(\T) $. There exist $ K,K',w_0 > 0 $ such that, 
for every $ w \in \,]0, w_0] $, 
for every function $ u \in \Lspace^1(\T;\R_+)$, 
for every $ x \in \R $, 
and for every $ n\in\N$,
one has
\begin{subequations}
\label{Potential theory}
\begin{align}
\tE\big( \nE^{w \sum_{k=1}^n h(X_{k-1}^x) B_k}\, u( X_n^x) \big) \;
&\leq\; 
\frac{K}{w\sqrt{n}} \int_\T u(y) \dif y 
\qquad 
(wn\ge \kappa, \, w^2 n \le 1)
\,, 
\label{upper bound}
\\
\tE \big( \nE^{w \sum_{k=1}^n h(X_{k-1}^x) B_k}\, u(X_n^x) \big) \;
&\ge\; 
K' \int_\T u(y) \dif y 
\qquad 
(1/2 \leq w^2n \leq 1)
\,.
\label{lower bound}
\end{align}
\end{subequations}
}

Before starting the proof let us make a few of definitions: First, for $ A \subset \T $ and $ 1 \leq p \leq \infty $ we define the space
\[
\Lspace^p_A(\T) \;:=\; \{ u\in\Lspace^p(\T):\spt(u)\subset A \}
\,.
\]
Secondly, let $ S $ be a continuous operator from $ \Lspace^p(\T) $ to $\Lspace^q(\T)$, for $ 1\leq p,q \leq \infty$, and denote the associated operator norm by $ \norm{S}_{p \to q } = \sup \{ \norm{Su}_q:u\in\Lspace^p(\T), \, \norm{u}_p \leq 1 \} $.

The content of Proposition \ref{prp:Potential theory} is twofold. 
First, it describes the approach to equilibrium of the chain $X$.
To see this, let us consider the case $h=0$, and let us take some subset $A\subset \T$.
Equation \eqref{upper bound} implies that $\Prob(X_n^x \in A) \lesssim \max \lbrace 1/(w\sqrt n), 1 \rbrace\mathrm{Leb}(A)$ when $wn\ge \kappa$, 
whereas \eqref{upper bound} and \eqref{lower bound} imply that $\Prob(X_n^x \in A)\sim \mathrm{Leb}(A)$ when $w^2n\ge 1/2$.
This is obvious when $w^2n \le 1$.
But, if $w^2n>1$, one can write $n=n_1 +n_2$ such that $\ceil{n_2 w^2} =1$, and
\[
\Expectation\msp{1}u(X_n^x) \;=\; 
\int_{\mathbb T } \Expectation \big( u(X_n^x)| X_{n_1}^x = y \big) \, \Prob (X_{n_1}^x \in\dif  y)
\,.
\]
The result follows since, if $ y \in \T$, one has $\Expectation \bigl[ u(X_n^x)\big| X_{n_1}^x = y \bigr] = \Expectation\msp{1}u(X_{n_2}^y) \sim \norm{u}_1 $.


Secondly, Proposition \ref{prp:Potential theory} asserts that the result obtained for $h=0$ is not destroyed when some specific perturbation is added ($h\ne 0$).
If $h\ne 0$ but if $u=1$, results \eqref{upper bound} and \eqref{lower bound} are trivial.
Indeed, by Azuma's inequality \eqref{Azuma's inequality}, one finds some $C>0$ such that, for every $n\in\N$ and for every $a>0$,
one has
\[
\Prob \Big( 
\nE^{-a} \le \nE^{w \sum_{k=1}^n h(X_{k-1}^x) B_k}  \le \nE^{a} \Big)
\;\ge\; 
1 \,-\,  2\msp{1} \nE^{-\frac{Ca^2}{w^2n}}
\,.
\]
So, in general, one sees that the rare events where 
$\nE^{w \sum_{k=1}^n h(X_{k-1}^x) B_k} $ is very large or very close to zero may essentially be neglected. 

In the sequel, one assumes that 
\begin{itemize}
\item[(A1)] $ \kappa>0$ and $h\in\Cont^1(\T)$ are given, 
\item[(A2)] $ w\in\,\rbrack 0, w_0\rbrack$, where $w_0$ is small enough to make all our assertions valid.
\end{itemize}
All the constants introduced below may depend on $\kappa$ and $h$.


In order to prove Proposition \ref{prp:Potential theory}, let us introduce a continuous operator $T$ on $\Lspace^p(\T)$, $1 \leq p\leq \infty $, by setting
\bels{T}{
T u (x) \;:=\; 
\Expectation \big[  (1 + wh(x)B)\, u\circ f_B(x) \big] 
\;=\;  
\int_{b_-}^{b_+} u\circ f_b(x) \, (1 + wh(x)b)\, \tau (b)\dif b
\,.
}
Since $\Expectation (B) = \int b \, \tau (b)\dif b = 0$, one has $T1 = 1$ and $\norm{T}_{\infty\to\infty } = 1$. 
The operator $T$ is thus, formally, the transition operator of some Markov chain on the circle.  
But, for every $b\in\lbrack b_-,b_+\rbrack$ and every $x\in\T$, one has
\[
\nE^{wh(x)b} \;=\; (1 + wh(x)b) \cdot \nE^{\mathcal O(w^2) }
\,.
\]
Therefore, for every $u\in\Lspace^1(\T;\R_+)$, for every $n\in\N$ satisfying $w^2 n\le 1$,
and for almost every $x\in\T$, one has
\bels{Ex Tn}{
T^n u (x) \;\sim\;  \Expectation\bigl( \nE^{w \sum_{k=1}^n h(X_{k-1}^x) B_k}\, u( X_n^x)\bigr)
\,.
}
Let $ y \in \T $. The proof of  Proposition \ref{prp:Potential theory} rests on the fact that, when $T^n$ acts on a function $u\in\Lspace^1_{\Ball(y,w^2)}(\T)$, it can be well approximated by an operator $S_{y,n}$ which can be explicitly studied.
In order to define $S_{y,n}$, let us first introduce the convolution operator $ T_y $ on $ \Lspace^p(\T) $, $1\le p\le \infty$, by setting
\bels{Ty}{
T_y u (x) := \int u\circ g_b(x,y) \,  (1 + wh(y)b) \, \tau (b)\dif b
\,,
}
where
\bels{gbxy}{
g_b(x,y) \;:=\; x \,+\, \vartheta \,+\, \Phi(y,b) \;=\; x \,+\, w \,+\, w\msp{1}\phi(y)b \,+\, w^2 \psi (y) b^2 \,+\, \mathcal O(w^3)
\,, 
}
with $ \Phi $ defined as in \eqref{def of Phi}, and $\phi$ and $\psi$ defined as in \eqref{def of phi} and \eqref{def of psi}.
Then, one sets $S_{y,0}:= \mathrm{Id}$, and defines each $ n \in \N $ 
\begin{subequations}
\begin{align}
\label{Syn}
S_{y,n} &:=\; T_{y-nw} \cdots  T_{y-w}.
\\
\label{Ry}
R_y &:=\; T - T_y.
\end{align}
\end{subequations}

The core of our approximation scheme is described by equation \eqref{approximation scheme} below, 
but let us now describe it heuristically. 
Let $z\in\T$, and let $u\in \Lspace^1_{\Ball(z,w^2) }(\T; \R_+)$. 
The support of $u\circ f_b$ should be centered at $z-w$, and so $g_b(\genarg, z-w)$ is likely to be the best approximation of $f_b$, among all the maps $g_b(\genarg, y)$ ($y\in\T$).
Therefore, one can think of $T_z$ as one of the best approximations of $T$ among all the operators $T_y$ ($y\in\T$).
One writes
\bels{loc heuristique 2}{
T^n u  \;=\; T^{n-1}\msp{-2}R_{z-w} u  \,+\, T^{n-1}T_{z-w} u
\,, 
}
where $R_{z-w}$ is defined by \eqref{Ry}. The first term in the right hand side of \eqref{loc heuristique 2} can be bounded by means of our estimates on $R_{y}$ ($y\in\T$), in Lemmas \ref{lmm:estimates Rk} or \ref{lmm:estimates Rk L1} below.
One is thus left with the second term. 
From the definition \eqref{Ty} of $T_y$ ($y\in\T$), the function $T_{z-w}u$ will be approximately centered at $z-w$. One now approximates $T$ by $T_{z-2w}$ and one obtains
\[
T^{n-1}T_{z-w} u \;=\; T^{n-2}R_{z-2w} T_{z-w} u \,+\, T^{n-2}T_{z-2w}T_{z-w} u
\,. 
\]
Again, one is left with the second term. 
But, continuing that way, one finally needs to handle the term $T T_{z-(n-1)w}\dots T_{z-w} u$, and one arrives to
\bels{loc heuristique 1}{
T T_{z-(n-1)w} \cdots T_z u \;=\; R_{z - nw}T_{z-(n-2)w} \cdots T_{z-w} u \;+\; T_{z-nw} \cdots T_{z-w} u
\,,
}
By the definition \eqref{Syn}, one has $T_{z-nw} \cdots T_{z-w} u = S_{z,n} u$. So, this time, the second term in \eqref{loc heuristique 1} can be bounded from above and below by some explicit estimates contained in Lemma \ref{lmm:approximate kernels} below. By means of Lemmas \ref{lmm:estimates Rk} and \ref{lmm:estimates Rk L1}, one thus needs to show that the sum of the terms containing an operator of the form $R_y$ ($y\in\T$) do not destroy the estimate on $S_{z,n}u$.

The rest of the section is organized as follows.
In Lemma \ref{lmm:approximate kernels}, one obtains some bounds on the functions $S_{y,n}u$ for $u\in\Lspace^1_{\Ball(y,w^2)}(\T)$.
The same bounds should be obtained for a Gaussian of variance $nw^2$ centered at $y$. 
The proof turns out to be a straightforward computation, since the operators $T_y$ are diagonal in Fourier space. 
Next, Lemmas \ref{lmm:estimates Rk} and \ref{lmm:estimates Rk L1} give us bounds on $R_y$.
Lemma  \ref{lmm:estimates Rk L1} is actually not crucial, and needs only to be used when $n < 8$, since then
the function $S_{y,n}u$ may not be smooth enough for Lemma \ref{lmm:estimates Rk} to be applied.
Some easy results about the localization of the functions $T^n u$ and $S_{y,n} u$, for $u\in\Lspace^1_{\Ball(y,Cw)}(\T)$, are then given in Lemma \ref{lmm:localization measures}.
Finally, the proof of Proposition \ref{prp:Potential theory} is given.

Let us notice that, in Lemma \ref{lmm:approximate kernels}, and consequently in the proof of Proposition \ref{prp:Potential theory}, 
one has to distinguish between the case where $y\sim 0$, and the case where $y$ is away from 0.
This comes from the lack of diffusivity of the chain $X$ around 0 (see property (iii) of Corollary \ref{crl:the three qualitative properties of X}).

\Lemma{approximate kernels}{
Let $\epsilon>0$.
There exists $K>0$ such that, 
for every $n\in\N$ satisfying $8 \le n \le w^{-2}$,
for every $y\in\T-\Ball(0,\epsilon)$, 
and for every $u\in\Lspace ^1_{\Ball(y,w^2)}(\T; \R_+)$, 
one has $S_{y,n} u \in \Cont^2(\T)$ and, for every $x\in\T$,
\begin{subequations}
\begin{align}
&\abs{\partial_x^l S_{y,n} u(x)} 
\;\leq\; 
\frac{\msp{-3}K\msp{1}\norm{u}_1}{(w\sqrt{n})^{(l+1)}}
\,, 
\qquad\quad l=0,1,2, 
\label{approximate kernels 1}
\\
&\absb{\sin^k \pi (x+wn-y) \cdot \partial_x^k S_{y,n} u (x)} 
\;\leq\; 
\frac{K \norm{u}_1}{w\sqrt{n}}\,, 
\quad k=1,2
\,. 
\label{approximate kernels 2}
\end{align}
\end{subequations}
Moreover, when $ \epsilon $ is small enough, 
there exists $K'(\epsilon) >0$, with $K'(\epsilon)\to \infty$ as $\epsilon\to 0$, such that, 
for every $n\in\N$ satisfying $\epsilon\le w^2n\le 2 \epsilon$, 
for every $x,y\in\T$,
and for every $u\in\Lspace ^1_{\Ball(y,w^2)}(\T; \R_+)$, 
\bels{approximate kernels 3}{
\abs{S_{y,n} u (x)} \;\ge\; K'(\epsilon)\norm{u}_1  
\quad\text{when}\quad \abs{x+nw-y}_\T \leq\, 10\msp{1} \epsilon
\,.
}
}
The proof is deferred to the Appendix \ref{assec:proof for approximate kernels}. 

\Lemma{estimates Rk}{
There exists $K>0$ such that, for every $u\in\Cont^2 (\T)$ and every $y\in \T$, one has
\bels{estimates Rk}{
\norm{R_y u}_\infty 
\,\leq\;  K w^2 \Bigl\{ \,&\normb{\sin \pi(\genarg - y-w) \cdot u'}_\infty 
+\, w \msp{1}\norm{u'}_\infty \\
+\, &\normb{\sin^2 \pi (\genarg - y-w) \,\cdot\, u''}_\infty  
+\, w \msp{1}\norm{u''}_\infty\Bigr\}
\,.
}
}
\begin{Proof}
One takes some $u\in\Cont^2(\T)$, and one fixes $x,y\in\T$. 
From the definitions \eqref{T} and \eqref{Ty}, one has 
\bea{
R_y u (x) \;\equiv\; (T- T_y) u (x)
 \;=&\; \int (u\circ f_b(x) - u\circ g_b(x,y))\, (1 + wh(x)b)\, \tau(b)\dif b
 \\
 &+\, w\msp{1}(h(x) - h(y))\int u \circ g_b (x,y)\, b \, \tau (b)\dif b
 \\
=:&\;  A_1 \,+\, A_2
\,.
}
It is enough to bound $|A_1|$ and $|A_2|$ by the right hand side of \eqref{estimates Rk}.

Let us first bound $ \abs{A_1} $. By the mean value theorem, and the definitions \eqref{expansion of fb, def of phi and psi} and \eqref{gbxy} of $ f_b $ and $ g_b $, one has
\[
u\circ f_b (x) \,-\, u\circ g_b (x,y) \;=\; 
u'(x + w + \xi_1)\, 
\Bigl( 
w \bigl[\phi(x)-\phi(y)\bigr]b \,+\, w^2 \bigl[\psi(x)-\psi(y)\bigr] b^2 \,+\, \mathcal O(w^3)
\Bigr)
\,,
\]
where $\xi_1 \equiv \xi_1(b)$ is such that
\begin{equation}
\label{bound-xi-1}
|\xi_1 | \;\leq\; w\msp{1}|\phi(x)-\phi(y)| \,+\, \mathcal O(w^2)
\,. 
\end{equation}
By the mean value theorem again, one has
\[
u'(x+w+\xi_1) \;=\; u'(x+w) \,+\, u''(x+w+\xi_2)\, \xi_1
\,, 
\]
where $ \xi_2 \equiv\xi_2(b)$ is such that $|\xi_2 |\le |\xi_1 |$.

Therefore, setting $\tilde\tau (b) = (1 + wh(x)b)\, \tau(b)$, one can write $A_1$ as  
\bea{
A_1 \;=\; u'(x+w)\int \Big( w \bigl[ \phi(x)-\phi(y)\bigr] b \,+\, w^2 \bigl[\psi(x)-\psi(y)\bigr] b^2 \,+\, \mathcal O(w^3)\Big) \tilde\tau(b)\dif b&
\\ 
+
\int u''(x+w+\xi_2(b))\, \xi_1(b)\,  \bigl( w \bigl[\phi(x)-\phi(y)\bigr] b \,+\, \mathcal O(w^2)\bigr) \tilde\tau(b)\dif b&
\,.
}
One has
\[
|\phi(x)-\phi(y) | 
\;\lesssim\; 
|\sin\pi(x-y)| 
\qquad \text{and}\qquad 
|\psi(x) - \psi(y) |
\;\lesssim\; 
|\sin\pi (x-y)|
\,.
\]
So, taking into account the bound \eqref{bound-xi-1} and the fact that $\int b\tau(b)\dif b = 0$, 
one gets
\bels{Rk loc 1}{
|A_1| 
\;\lesssim\;    
&w^2 |u'(x+w)| \,  |\sin \pi(x-y)| \,+\, w^3 \norm{u'}_\infty 
\\
+\; &w^2 \!\int |u''(x+w+\xi_2(b))|\,\sin^2\pi(x-y) \, \tilde\tau (b)\dif b \,+\, w^3 \norm{u''}_\infty
\,. 
} 
But one has $\sin^2\pi (x-y)  \le  \sin^2 \pi (x + \xi_2-y) + \mathcal O(w)$.
So, inserting this last bound in \eqref{Rk loc 1}, one sees that $|A_1|$ is bounded by the right hand side of \eqref{estimates Rk}.

Let us then bound $|A_2|$.
By the mean value theorem and the definition \eqref{gbxy} of $g_b$, one writes
\[
u\circ g_b(x,y) \;=\; u(x + w) \,+\, u'(x+w + \xi)\, \mathcal{O}(w)
\,, 
\]
where $\xi \equiv \xi (b) = \mathcal O(w)$. Therefore, taking into account that $\int b\tau(b)\dif b=0$ and that
$ |h(x) - h(y)| \lesssim |\sin\pi(x-y)| $, one obtains
\bea{
|A_2| \;&\lesssim\; w^2  \int |\sin\pi(x-y)|\cdot |u'(x+w+\xi(b))|\cdot |b| \tau (b)\msp{1}\dif b   
\\
&\lesssim\; 
w^2 \bigl(\, \norm{\sin \pi(\mathrm {Id} - y-w) \cdot u'}_\infty +\, w\norm{u'}_\infty \bigr)
\,.
}
This finishes the proof.
\end{Proof}

\Lemma{estimates Rk L1}{
Let $ K,\epsilon >0 $. Let $ y\in\T $ be such that $ \abs{y}_\T \ge \epsilon $. 
Then there exists $ K'>0 $ such that, for every $u\in \Lspace^1_{\Ball(y,Kw) }(\T)$, one has
\bels{estimates Rk L1}{
\norm{R_y u}_1 \le  K' w \norm{u}_1
\,. 
}
Moreover $ Tu \in \Lspace^\infty(\T) $, and one has
\bels{estimates T L1}{
\norm{Tu}_\infty \le K' \, w^{-1}\, \norm{u}_1
\,.
}
}
\begin{Proof}
The constants introduced in this proof may depend on $K$ and $\epsilon$.
Let $u\in \Lspace^1_{\Ball(y,Kw) }(\T)$. 
One writes
\bels{T Ty noyaux}{
Tu(x) = \int_{\Ball(y,Kw) } t(x,z) u(z)\dif z 
\quad\text{and}\quad 
T_yu(x) = \int_{\Ball(y,Kw)} t_y(x,z) u(z)\dif  z
\,, 
}
where the functions $t$ and $t_y$ are obtained by performing a change of variables in the definitions \eqref{T} and \eqref{Ty} of $T$ and $T_y$.
Setting $F_x(b):=f_b(x)$ and $G_x(b):=g_b(x,y)$, where  $f_b$ and $g_b$ are defined in \eqref{expansion of fb, def of phi and psi} and \eqref{gbxy},
one obtains
\bels{pot th l1 2}{
t(x,z) \;&=\; (1 + w h(x)F_x^{-1}(z))\, \tau(F_x^{-1}(z))\, \partial_z F_x^{-1}(z)
\,,
\\
t_y(x,z) \;&=\; (1 + w h(y)G_x^{-1}(z))\, \tau(G_x^{-1}(z))\, \partial_z G_x^{-1}(z)
\,.
}
Let $z\in \Ball(y,Kw)$ be given. 
Let us see that $t(\genarg, z)$ and $t_y(\genarg, z)$ are well defined functions.  The support of $t(\genarg, z)$ (respectively of $t_y(\genarg,z)$) is the support of $\tau\circ  F_{(\genarg)}^{-1}(z)$ (resp. of $\tau \circ G_{(\genarg)}^{-1}(z)$). The support of $\tau\circ  F_{(\genarg)}^{-1}(z)$ is made of all the $x$ such that 
\[
b_- \le F_x^{-1}(z)\le b_+ \; \Leftrightarrow\;  f_{b_-}(x) \le z \le f_{b_+}(x)\;  \Leftrightarrow\;  f_{b_+}^{-1}(z) \le x \le f_{b_-}^{-1}(z)
\,. 
\]
One obtains a similar relation for the support of $\tau \circ G_{(\genarg)}^{-1}(z)$ and one gets therefore 
\[
\spt(t(\genarg,z))\,, \; \spt(t_y(\genarg , z)) \,\subset\, \Ball (z, Cw) \,\subset\, \Ball(y, C'w).
\]
The hypothesis $|y|_\T \ge \epsilon$ ensures that the maps $F_x$ and $G_x$ are invertible when $x\in \Ball(y, C'w)$, and actually that
\bels{pot th l1 1}{
\partial_b F_x(b) \,\gtrsim\, w 
\quad\text{and}\quad \partial_b 
G_x(b) \,\gtrsim\,  w
\,.
}
This shows in particular that  $t(\genarg, z)$ and $t_y(\genarg, z)$ are bounded functions.

Let us now show \eqref{estimates Rk L1}.
Taking \eqref{pot th l1 2} into account, one has, from the definition \eqref{Ry} of $R_y$, 
\bels{}{
\norm{R_y u}_1 \le \int_{\Ball(y,Kw) } |u(z)|\dif z \int_{\Ball(y,C'w) }|t(x,z) - t_y(x,z)|\dif x.
}
It is therefore enough to show that, for every $z\in \Ball(y,Kw)$, one has
\bels{pot th l1 a}{
\int_{\Ball(y,C'w) }|t(x,z) - t_y(x,z)|\dif x \;=\; \mathcal O(w)
\,.
}

Let us take some  $z\in \Ball(y,Kw)$ and some $x\in\Ball(y,C'w)$.
Since $b_-\le F_x^{-1}(z),G_x^{-1}(z)\le b_+$, since $\tau$ is bounded, and since \eqref{pot th l1 1} holds, one finds, starting from  \eqref{pot th l1 2},  that
\bels{pot th l1 b}{
|t(x,z) - t_y(x,z)| 
\;\lesssim\;
|\partial_z F_x^{-1}(z) - \partial_z G_x^{-1}(z) | \,+ w^{-1} |\tau (F_x^{-1}(z)) - \tau (G_x^{-1}(z)) | \,+\, C.
}
For every $b\in\lbrack b_-,b_+\rbrack$, one has $\partial_b F_x(b) = w\phi(x) + \mathcal O(w^2)$ and $\partial_b G_x(b) = w\phi(y) + \mathcal O(w^2)$.
Therefore
\bels{pot th l1 c}{
|\partial_z F_x^{-1}(z) - \partial_z G_x^{-1}(z) | \;&\leq\; \Big|\frac{1}{w\phi (x) + \mathcal O(w^2) } - \frac{1}{w\phi (y) + \mathcal O(w^2) } \Big| 
\\
&\lesssim\; w^{-1}  |\phi (y) - \phi (x) + \mathcal O(w) | \;\lesssim\; 1
\,, 
}
since $|y - x| = \mathcal O(w)$.
Inserting thus \eqref{pot th l1 c} in \eqref{pot th l1 b}, and then \eqref{pot th l1 b} in \eqref{pot th l1 a}, one finds
\bels{pot th l1 z}{
\int_{\Ball(y,Cw) }|t(x,z) - t_y(x,z)|\dif x\;  
&\lesssim\; 
w^{-1} \int_{\Ball(y,Cw) } |\tau (F_x^{-1}(z)) - \tau (G_x^{-1}(z)) | \dif x \,+\, \mathcal O(w)
\\
&=: \;w^{-1} I \,+\, \mathcal O(w)
\,.
}
It remains thus to show that $I = \mathcal O (w^2)$.
For this, let us define
\[
D_1 \,:=\; \lbrace x\in\T : b_-\le F_x^{-1}(z)\le b_+ \rbrace\,, 
\quad\text{and}\quad 
D_2 \,:=\;  \lbrace x\in\T : b_-\le G_x^{-1}(z) \le b_+ \rbrace
\,.
\]
One writes 
\[
I \;= \int_{D_1\cap D_2 } (\dots ) + \int_{(D_1\cap D_2)^c}(\dots) \;=:\, I_1 \,+\, I_2 
\,.
\]
First, when $x\in D_1 \cap D_2$, one uses the fact that $\tau\in\Cont^1(\lbrack b_-,b_+\rbrack)$, that 
\[
|F_x^{-1}(z)-G_x^{-1}(z)| 
\;=\; 
\Big| \frac{z-x-w}{w\phi(x) }  - \frac{z-x-w}{w\phi(y)} + \mathcal O(w)\Big| 
\;=\; 
\mathcal O(w)
\,,
\]
since $|z-x-w|=\mathcal O(w)$ and $|\phi (y)-\phi(x) |=\mathcal O(w)$, 
and that $\mathrm{Leb}(D_1\cap D_2)= \mathcal O(w)$, to conclude that $I_1=\mathcal O(w^2)$.
Next, when $x\in (D_1\cap D_2)^c$, one has $t(x,z)=t_y(x,z)=0$, except on $D_1\, \Delta\, D_2$.
But, for every $b\in\lbrack b_-,b_+\rbrack$, one has $|f_b(x)-g_b(x,y)| = \mathcal O(w^2)$, since $|x-y|=\mathcal O(w)$.
So, one has $\mathrm{Leb}(D_1\, \Delta\, D_2) = \mathcal O(w^2)$, and thus $I_2= \mathcal O(w^2)$.

Let us finally show \eqref{estimates T L1}.
From \eqref{T Ty noyaux}, one has that $|Tu(x)| \le \mathrm{sup}_{z\in\Ball(y,Kw) }|t(x,z)|$. 
The relations \eqref{pot th l1 2} and \eqref{pot th l1 1} allow us to obtain the result. 
\end{Proof}

In order to prove the next lemma, we introduce the adjoint $T^*$ of $T$ with respect to the Lebesgue measure.
This operator is defined on $\Lspace^p(\T)$ ($1\le p \le \infty$) and is
such that, 
for every $u\in\Lspace^p(\T)$ and every $v\in\Lspace^{p'}(\T)$, with $1/p + 1/p' = 1$, 
one has
\bels{def adjoint}{
\int_{\T} v\, T^* u\, \dif x = \int_{\T} u\,  Tv\, \dif x
\,.
} 
From the definition \eqref{T} of $T$, one concludes that
\bels{adjoint}{
T^* u (x) \;= \int_{b_-}^{b_+} u\circ f_b^{-1}(x)\, \bigl[1 + w \, h\circ f_b^{-1}(x)\msp{1}b\bigr]\, \partial_x f_b^{-1}(x)\, \tau (b)\dif b
\,.
}
Therefore, when $u\ge 0$, one has
\bels{approximation adjoint}{
T^* u (x) \;\ge\; 
\nE^{-\mathcal O(w)} \int u\circ f_b^{-1}(x)\, \tau (b)\dif b
\,. 
}
For $z\in\R$, let us define the chain $Y =(Y_{n}^z : n\in\N_0)$ by $Y_{0}^z := z $ and 
\bels{chaine Yzn}{
Y_{n}^z \;:=\; f_{B_n}^{-1}(Y_{n-1}^z) \;=\; Y_{n-1}^z -\, w \,-\,  w \phi (Y_{n-1}^z)B_n \,+\, \mathcal{O}(w^2)
\,.
}

\Lemma{localization measures}{
Let $K>0$. There exist $K_2 \ge K_1>0$ such that, for every $n\in\N$, for every $y\in\T$, 
and for every $u\in\Lspace^1_{\Ball(y,Kw)}(\T)$, one has
\bels{support measures}{
\spt(T^n u),\; \spt(S_{y,n}u) \,\subset\, \bigl[ y- K_2wn, y- K_1 wn \bigr]
\,.
}

Morover, for every $ R>0 $ large enough, there exists $K'>0$ such that,  for every $n\in\N$ satisfying $wn \leq 1 $, 
for every $y\in\T$,  and for every $u\in \Lspace^1_{\Ball(y,w)}(\T;\R_+)$, one has
\bels{concentration measure}{
\int_{\Ball(y-nw, R\sqrt w)} T^n u (z) \dif z  \;\ge\; K' \norm{u}_1
\,.
}
}
\begin{Proof}
Let us first show \eqref{support measures}.
Let us consider the case of $T^n u$ ; the case of $S_{y,n}u$ is strictly analogous.
From the definition \eqref{T}, one sees that
\[
\spt(T^n u ) 
\;\subset\; \bigl[ f_{b_+}^{-n}(y-Kw/2), f_{b_-}^{-n} (y+Kw/2) \bigr]
\,.
\]
This implies the result, since, by the definition \eqref{expansion of fb, def of phi and psi} of $f_b$, one has, for every $x\in\T$ and every $b\in\lbrack b_-,b_+ \rbrack$,
\[
(1+b_-)w - \mathcal O(w^2) 
\;\leq\; 
x - f_b^{-1}(x) 
\;\leq\; 
(1+b_+)w + \mathcal O(w^2)
\,.
\]
Let us then show \eqref{concentration measure}.
Let $u\in\Lspace^1_{\Ball(y,w^2)}(\T;\R_+)$, let $R>0$, and let $n\in\N$ be such that $nw \le 1$.
From the definition \eqref{def adjoint} of the adjoint $T^*$, one has 
\[
\int_{\Ball(y-nw, R\sqrt w) } T^n u (z) \dif z \;=\; \int_{\Ball(y,w) } T^{*n}\chi_{\Ball(y-nw,R\sqrt w) }(z) 
\, u (z)\dif z
\,.
\]
It is therefore enough to show that, for every $z\in \Ball(y,w)$, one has $T^{*n}\chi_{\Ball(y-nw,R\sqrt w) }(z) \gtrsim 1$, if $R$ is large enough.
But, since $wn\le 1$, \eqref{approximation adjoint} implies that
\bels{pot th locc 2}{
T^{*n}\chi_{\Ball(y-nw,R\sqrt w) }(z) \;
&\gtrsim\; 
\Expectation\bigl(\chi_{\Ball(y-nw,R\sqrt w)}\circ f_{B_n}^{-1}\circ \dots \circ f_{B_1}^{-1}(z) \bigr)
\\
&=\;  1 \,-\; \Prob\bigl( \abs{Y_{n}^z - (y-nw)}   \ge R\sqrt{w}\bigr)
\,,
}
where $Y$ is defined in \eqref{chaine Yzn}.
Therefore, since $|z-y|=\mathcal O(w)$ and since $w^2n = \mathcal O(w)$, 
one obtains, from the definition \eqref{chaine Yzn} of $Y$, and from Azuma's inequality \eqref{Azuma's inequality}, that
\bels{pot th locc 1}{
\Prob\bigl( \abs{Y_{n}^z - (y-nw)} \ge R \sqrt{w}\bigr) \;=\; \Prob\Big(  \Big| w\sum_{k=1}^n \phi (Y_{k-1}^z) + \mathcal O(w)\Big| \ge R\sqrt w \Big) 
\;\leq\; 2 \msp{1} \nE^{-\frac{CR^2}{nw}} 
\,.
}
The proof is finished by taking $R$ large enough, and inserting \eqref{pot th locc 1} in \eqref{pot th locc 2}.
\end{Proof}

\begin{Proof}[Proof of proposition \ref{prp:Potential theory}]
Let $n\ge 9$ be such that $nw^2 \le 1$.
Let us make three observations.
First, by \eqref{Ex Tn}, 
it is enough to show the proposition with $\Expectation_x (\nE^{w \sum_{k=1}^n h(X_{k-1}) B_k}\, u( X_n))$ replaced by $T^n u(x)$ in \eqref{upper bound} and \eqref{lower bound}. 

Second, it is enough to prove the proposition for functions in $\Lspace^1_{\Ball(y,w^2) }(\T; \R_+)$ for every  $y\in\T$.
So, throughout the proof, one assumes that $y\in\T$ is given, and the symbol $ v $ denotes a function in $\Lspace^1_{\Ball(y,w^2) }(\T; \R_+)$.

Third, it is enough to show \eqref{lower bound} for some $n'$ satisfying $w^2 n' \le 1/2$. 
Indeed, let us now assume that \eqref{lower bound} is shown for this $n'$, and let $n$ be such that $1/2 \le w^2 n \le 1$. 
From the definition \eqref{T}, one sees that, if $u_1\ge u_2$, one has $Tu_1\ge Tu_2$.
So, one writes $n = n' + n''$ and, for every $u\in\Lspace^1(\T,\R_+)$, one gets 
$T^n u (x) = T^{n''}T^{n'}u \gtrsim \norm{u}_1 T^{n''}1 \sim \norm{u}_1$, 
where the fact that $T^{n''}1 \sim 1$ directly follows from the definition \eqref{T} of $T$, Azuma's bound \eqref{Freedman's and Azuma's exponent bounds}, and the hypothesis $ w^2 n\le 1 $.

The proof is now divided into three steps, but the core is entirely contained in the first one.

\textbf{Step 1: approximating $T^n$ by $S_{y,n}$:}
One here shows the bounds \eqref{upper bound} and \eqref{lower bound} under two particular assumptions:
\begin{enumerate}
\item One supposes that $|y|_\T \ge \epsilon_1$, for some $\epsilon_1>0$. The constants introduced below may depend on $\epsilon_1$.
\item Only for \eqref{lower bound}, 
one assumes that $n$ is such that $\epsilon_2\le n \le 2 \epsilon_2$ and that $|x+nw-y|_\T \le 10 \epsilon_2$ for some $\epsilon_2>0$ small enough.
\end{enumerate}
By the definition \eqref{Syn} of $S_{y,n}$, one can write
\bels{approximation scheme}{
T^n v \;&=\; S_{y,n}v \,+\, \sum_{k=1}^8 T^{n-k}R_{y-kw}S_{y,k-1}v \,+\,  \sum_{k=9}^{n-1} T^{n-k}R_{y-kw}S_{y,k-1}v 
\\
&=:\; S_{y,n}v \,+\, Q_1 \,+\, Q_2
\,.
}

Let us bound $\norm{Q_1}_\infty$.
Let $k\in\N$ be such that $ 1\le k\le8 $.
By \eqref{support measures}, one has
\bels{pot th loc abc}{
\spt(R_{y-kw}S_{y,k-1}v) \,\subset\, \Ball(y,Cw)
\,.
}
Remembering that $\norm{T}_{\infty\to\infty}=1$, one uses \eqref{estimates Rk L1} and \eqref{estimates T L1} to obtain that 
\bels{pot th S1}{
\norm{Q_1}_\infty &\leq\, \sum_{k=1}^8 \norm{T^{9-k}R_{y-kw}S_{y,k-1}v}_\infty 
\;\lesssim\; 
w^{-1} \sum_{k=1}^8 \norm{R_{y-kw}S_{y,k-1}v}_1
\\
&\lesssim\, \sum_{k=0}^7 \norm{ S_{y,k-1}v}_1 \;\lesssim\; \norm{v}_1
\,,
}
where, for the last inequality, one has used the fact that $\norm{T_y}_{1\to 1 } = 1$ for every $y\in\T$. 

Let us  bound $\norm{ Q_2}_\infty$. By Lemma \ref{lmm:estimates Rk} and estimates  \eqref{approximate kernels 2} and \eqref{approximate kernels 1} in Lemma \ref{lmm:approximate kernels}, one has, for $8\le k\le  w^{-2}$, 
\[
\norm{ T^{n-k}R_{y-kw}S_{y,k-1}v}_\infty 
\;\leq\; 
\norm{R_{y-kw}S_{y,k-1}v }_\infty 
\;\lesssim\; 
w^2 \biggl\{ \frac{1}{w\sqrt k } + \frac{w}{w^2k} +\frac{1}{w\sqrt k } +\frac{w}{w^3 k^{3/2}} \biggr\} \cdot \norm{v}_1.
\]
Therefore, since $w^2n \le 1$ by hypothesis, one gets
\bels{pot th S2}{
\norm{Q_2}_\infty 
\;\lesssim\; 
( w\sqrt n + w\log n + C)\norm{v}_1 \;\lesssim\; \norm{v}_1
\,.
}
So, from \eqref{approximation scheme},  \eqref{pot th S1} and \eqref{pot th S2},  
one has 
\[
\norm{T^n v  - S_{y,n}v}_\infty \;\leq\; C\, \norm{v}_1
\,,
\]
where the constant $C$ is independent of $\epsilon_2$.
Therefore, in the particular case considered, \eqref{upper bound} follows from \eqref{approximate kernels 1} with $l=0$,
and \eqref{lower bound} follows from \eqref{approximate kernels 3}, if $\epsilon_2$ has been chosen small enough.

\textbf{Step 2: proof of \eqref{upper bound}:}
By Step 1, \eqref{upper bound} is known to hold when $|y|_\T \ge \epsilon_1$, 
and one may now assume that $|y|_\T < \epsilon_1$.
Moreover, one has still the freedom to take $\epsilon_1$ as small as we want.
One now uses the hypothesis $nw \ge \kappa$.
Let $m\in\N$ be such that $mw = \epsilon'$, for some $\epsilon'\in \rbrack 0,c/2\rbrack$.
If $\epsilon_1$ is small enough,  
it follows from \eqref{support measures} that one can chose $\epsilon'$ such that $\mathrm{supp}(T^m u )\cap \Ball(0,\epsilon_1)=\emptyset$.
But
the particular case considered in Step 1 implies that \eqref{upper bound} is valid for any function in $\Lspace^1_{\T - \Ball(0,\epsilon_1)}(\T)$, and thus one has
\[
T^n v (x) \;=\; T^{n-m}T^m v (x) 
\;\lesssim\; 
\frac{\norm{T^m v }_1}{w\sqrt{n-m} } 
\;\lesssim\; 
\frac{\norm{v}_1}{w\sqrt{n}}
\,,
\]
where the last inequality follows from the fact that $\norm{T}_{1\to 1} \le \nE^{\mathcal O(w) }$, as can be seen from the definition \eqref{T}.

\textbf{Step 3: proof of \eqref{lower bound}:} 
One first will establish \eqref{lower bound} for $n$ such that $n=\lfloor \epsilon_2w^{-2} \rfloor$, and for $x$ such that $|x+nw-y|_\T \le 10 \epsilon_2$.
By Step 1, it is now enough to consider the case $|y|_\T < \epsilon_1$.
Let now $m = \lfloor  \frac{1}{2}w^{-1}\rfloor$, and let $R>0$.
If $R$ is taken large enough,  
it follows from \eqref{concentration measure}, 
and from the particular case of \eqref{lower bound} already established in Step 1, 
that
\[
T^n v (x) \,\ge\, T^{n-m} \msp{-1}\big( \chi_{\Ball(y-mw,R\sqrt w)} T^m v \big)(x) 
\;\gtrsim\; 
\int_{\Ball(y-mw,R\sqrt w)} T^m v (z)\dif z 
\;\gtrsim\; 
\norm{v}_1
\,.
\]

One finally needs to get rid of the assumption $|x+nw-y|_\T \le 10 \epsilon_2$.
One uses a classical technique \cite{Coulhon-1993}.
One shows \eqref{lower bound} for $n=kq$, with $k\ge 1/ 18 \epsilon_2$, and $q$ such that $q = \lfloor \epsilon_2 w^{-2}  \rfloor$.
One already knows that 
\bels{pot th: Tq v}{
T^{q}v 
\;\gtrsim\;  
\chi_{\Ball(y-qw,10 \epsilon_2) }\norm{v}_1
\,. 
}
But one now will show that, for every $z\in\T$, and for every $s\in\lbrack \epsilon_2, 1\rbrack$,
one has 
\bels{pot th: Tq chi}{
T^q  \chi_{\Ball(z,s)} \gtrsim \epsilon_2\,  \chi_{\Ball(z-qw,s+9\epsilon_2) }. 
}
This will imply the result : 
\bea{
T^{n}v \;&=\; T^{kq}v 
\;\gtrsim\;  
T^{(k-1)q} \chi_{\Ball(y-qw,10 \epsilon_2) } \, \norm{v}_1 
\\
&\gtrsim\, \dots \,\gtrsim\;  
\epsilon_2^{k-1}\chi_{\Ball(y-kqw, (10 + 9(k-1))\epsilon_2) } \, \norm{v}_1 
\;\gtrsim\;  
\epsilon_2^{k-1} \norm{v}_1
\,.
}

Let us thus show \eqref{pot th: Tq chi}.
Let $z\in\T$ and $s\in\lbrack \epsilon_2, 1\rbrack$. 
Let us write $T^q u (x) = \int t_q (x,z')u(z')\dif z'$ for any $u\in\Lspace^1(\T)$. 
Relation \eqref{pot th: Tq v} implies in fact that $t_q (x, \genarg ) \gtrsim \chi_{\Ball(x+qw,10 \epsilon_2) }(\genarg)$
(which may be formally checked by taking $u(x) = \delta (y-x)$). 
Therefore
\bea{
T^q \chi_{\Ball(z,s)}(x)\;
&\gtrsim\;   \int \chi_{\Ball(x + qw,10\epsilon_2) }(z') \cdot \chi_{\Ball(z,s) }(z') \dif z' 
\\
&\gtrsim\; \epsilon_2\, \chi_{\Ball(z,s+9\epsilon_2) }(x+qw) 
\;=\; \epsilon_2\, \chi_{\Ball(z-qw,s+9\epsilon_2) }(x)
\,.
}
This finishes the proof.
\end{Proof}

\section{Putting everything together}
\label{sec:Proof of Theorem}

In \cite{Casher-Lebowitz-71} p. 1710, Casher and Lebowitz derive the lower bound $\tE(J_n) \gtrsim (T_1 - T_n) n^{-3/2}$. However, their argument contains a gap, and consequently this lower bound remains still to be proven. Indeed, their proof is based on the estimate on the following estimate of $ D_n(e_1) $ ($ K_{1,n} $ in their notation):
\bels{cas leb estimate}{
\tE\bigl[ D_n(e_1)^2\bigr] \;\sim\; \nE^{Cnw^2} \quad\text{as}\quad w \searrow 0
\,.
}
This bound is obtained by computing the eigenvalues of a $ 4 \times 4 $ matrix $ F $, defined in \cite{Casher-Lebowitz-71} p. 1710.
But this estimate  cannot hold. 
Indeed, we know for example, from Corollary \ref{crl:Fundamental decompositions of D_1n and D_2n} and Proposition \ref{prp:Potential theory}, that $\Expectation(D_{1,n}^2)\sim w^{-2}$ when $w^2n \sim 1 $. 
Although the computation of the eigenvalues of $ F $ is correct, the authors do not take into account the fact that a $ w $-dependent change of variables is needed to obtain a correct estimate on $\Expectation[D_n(e_1)^2]$.

\subsection{Proof of the lower bound}
\label{ssec:Proof of the lower bound}

We begin by a lemma. Let $(L_n)$ and $(K_n)$ be the processes defined in Lemma \ref{lmm:Representations for the lower bound}.

\Lemma{bound for K_n and L_n probabilities}{
For every $\alpha >0$, there exists $C(\alpha)>0$, such that, for every $ a>0 $, and every $ n \in \N $ satisfying $ w^2n \leq 1 $, one has
\bels{bound for K_n and L_n probabilities}{
\tP(\abs{K_n} \ge a),\, \tP (\abs{L_n} \ge a) \;\leq\;  C(\alpha) w^\alpha
\,.
}
}
\begin{Proof}
Let $ (A_n:n\in\N_0)$ be a $\mathbb F$-adapted process such that 
\bels{def of A_n}{
A_n \;:=\; \nE^{M_n +\, L_n +\, \mathcal{O}(w^2n) } 
}
for every $n \in \N_0 $, with $ M_n $ as defined in Lemma  \ref{lmm:Representations for the lower bound}. From the expressions \eqref{def of dL_n} and \eqref{def of dK_n}, both $ K_n $ and $ L_n $ are of the form
\bea{
R_n \;:=\; w^2 \sum_{j=1}^n A_{j-1}S_{j-1}B_j
\,, 
}
where $ (S_j) $ is $\mathbb{F}$-adapted, and satisfies $ \abs{S_j} \lesssim 1 $ for $ j\in \N_0 $. 

Let $a>0$. One writes
\bea{
\tP (R_n \ge a) \;=\;
\tP\biggl( &w^{3/2} \sum_{j=1}^n w^{1/2} A_{j-1}S_{j-1}B_j \ge a, \; \max_{1\leq j\leq n}w^{1/2}\!A_{j-1} \leq 1\biggr) 
\\
+\;\tP\biggl( &w^{3/2} \sum_{j=1}^n w^{1/2} A_{j-1}S_{j-1}B_j \ge a, \; \max_{1\leq j\leq n}w^{1/2}\!A_{j-1}  > 1 \biggr)
\,.
}
Let us now define a process $ (\tilde A_n :n\in\N_0)$ by setting $\tilde{A}_n := A_n \cdot \chi_{[0,1]}\msp{-1}(w^{1/2}A_n) $. 
One has
\bels{final proof lemm loc 01}{
\tP( R_n \ge a ) 
\;\leq\; 
\tP\Big( w^{3/2} \sum_{j=1}^n w^{1/2}\tilde{A}_{j-1}S_{j-1}B_j \ge a \Big) 
\,+\, 
\sum_{j=1}^n \tP(w^{1/2}A_{j-1}> 1)
\,.
}
First, by Azuma's inequality \eqref{Azuma's inequality}, and since $ w^2n \leq 1$, one has
\bels{final proof lemm loc 02}{
\tP \biggl(w^{3/2} \sum_{j=1}^n w^{1/2}\tilde{A}_{j-1}S_{j-1}B_j \ge a \biggr) 
\;\leq\; 
2 \, \nE^{- Ca^2 /w^3n} 
\;\leq\; 
\nE^{- Ca^2w^{-1}}
\,.
}
Next, it follows from \eqref{def of dM_n}, \eqref{def of dL_n} and \eqref{increments bounded by constant} that $ A_n $ defined in \eqref{def of A_n} if also of the form $ A_n = \nE^{w\sum_{j=1}^n G_{j-1} B_j + \mathcal{O}(w^2 n)} $, where $(G_j) $ is $\mathbb{F}$-adapted, and $ \abs{G_j} \lesssim 1 $ for $ j \in \N_0 $.
So, applying again Azuma's inequality, one gets
\[
\tP(w^{1/2} A_{j-1}>1) \;=\; \tP\Big(w\sum_k^{j-1}G_{k-1}B_k + \mathcal{O}(w^2 n) >  \frac{1}{2}\log \frac{1}{w}\Big)
\;\lesssim\; 
\nE^{-\frac{ C \log^2 (1/w) }{(j-1)w^2}} 
\;\leq\; 
\nE^{- C' \log^2 (1/w)}
\,.
\]
Therefore $ \sum_{j=1}^n \tP(w^{1/2}A_{j-1}> 1) \lesssim w^{-2} \nE^{-C \log^2 (w^{-1}) } $. The proof is finished by inserting this last bound and \eqref{final proof lemm loc 02}  in \eqref{final proof lemm loc 01}.
\end{Proof}

With the help of this lemma we can now prove the lower bound $ \tE\msp{1}J^\mathrm{CL}_n \gtrsim n^{-3/2} $ of Theorem \ref{thr:the scaling of the average current}.
Indeed, from \eqref{CL-current and def of J_n}, it follows that
\bea{
\tE\msp{1}J^\text{CL}_n \;\gtrsim\;  \int_{(2n)^{-1/2}}^{n^{-1/2}} \tE\msp{1} j_n (w) \msp{1}\dif w
\,, 
}
with $j_n$ defined in \eqref{def of jn}. It is therefore enough to show that when $ 1/2 \leq w^2n \leq 1 $ the bound $ \Expectation j_n(w) \gtrsim w^2 \sim n^{-1} $ holds. 
So let $ 1/2 \leq w^2n \leq 1 $, and use Corollary \ref{crl:Fundamental decompositions of D_1n and D_2n} in \eqref{def of jn} to write
\bels{final proof: jn lower bound}{
j_n(w) 
\;\gtrsim\; 
\Bigg\{
1 \,+\,
\frac{(\Gamma_{n}^\vartheta\sin X_{n}^\vartheta)^2 }{w^4} + \frac{( \Gamma^\vartheta_{n-1} \sin X_{n-1}^\vartheta)^2 }{w^2}  + \frac{(\Gamma_n^0\sin X_{n}^0)^2 }{w^2}+ (\Gamma_{n-1}^0\sin X_{n-1}^0)^2
\Biggr\}^{-1}
\msp{-10}.
}

Let us take some $ R, c>1 $. The constants introduced below may depend on $R$ and $c$.
Let us observe that, by point (i) of Corollary \ref{crl:the three qualitative properties of X}, one has $ \abs{X_{n-1}}_\T \lesssim  w $ provided $ \abs{X_n}_\T \lesssim w^2 $, and that, from the definition \eqref{def of Gamma^x_n}, one has $\Gamma_{n-1}\in\lbrack 0,2R\rbrack$ when  $\Gamma_n \in\lbrack 0,R\rbrack$. It follows therefore from \eqref{final proof: jn lower bound} that
\bels{final proof: jn proba 1}{
\Expectation\msp{1}j_n(w) 
\;\gtrsim\;  
\Prob\bigl(
\abs{X_n^\vartheta}_\T\le w^2, \, \Gamma_n^\vartheta \le R, \, \abs{X_n^0}_\T \le cw, \, \Gamma_n^0 \leq cR
\bigr)
\,.
}
We now uses Lemma \ref{lmm:Representations for the lower bound}. 
First, by \eqref{R-distance between X^vartheta_n and X^0_n}, one has
\bels{final proof: with Mn and Ln}{
\chi_{\Ball(cw) }(X_n^0) \, \ge\,   \chi_{\lbrack 0, R\rbrack  }(\nE^{M_n}) \cdot \chi_{\Ball(0,1) }(L_n) \cdot \chi_{\Ball(0,w^2) }(X_n^\vartheta)
\,,
}
provided $ c $ is large enough. 
Secondly, by \eqref{Gamma_2n from Gamma_1n}, one has
\bels{final proof: with Kn}{
\chi_{\lbrack 0, cR\rbrack }(\Gamma_n^0) 
\;\ge\;  
\chi_{\Ball(0,1) }(K_n) \cdot  \chi_{\lbrack 0,R\rbrack }(\Gamma_n^\vartheta)
\,,
}
again, provided $ c $ is large enough.
Using then  \eqref{final proof: with Mn and Ln} and \eqref{final proof: with Kn} in \eqref{final proof: jn proba 1}, one obtains
\bea{
\Expectation\msp{1}j_n(w) 
\;\gtrsim\; 
&\Prob( 
|X_n^\vartheta |_\T\le w^2, \, \Gamma_n^\vartheta \le R, \, \nE^{M_n} \le R, \, |L_n|\le 1, \, |K_n| \le 1)
\\
\;\ge\;
&\Prob(|X_n^\vartheta |_\T\le w^2, \, |L_n|\le 1, \, |K_n| \le 1)
\\
&-
\Prob(|X_n^\vartheta |_\T\le w^2, \,  \Gamma_n^\vartheta > R)
\,-\,
\Prob(|X_n^\vartheta |_\T\le w^2, \, \nE^{M_n} > R) 
\\
\; \ge &\; 
\Prob(|X_n^\vartheta |_\T\le w^2)
\,-\, \Prob(|L_n| > 1) - \Prob(|K_n| > 1)
\\
&-\
\Prob(|X_n^\vartheta |_\T\le w^2, \,  \Gamma_n^\vartheta > R)
\,-\,
\Prob(|X_n^\vartheta |_\T\le w^2, \, \nE^{M_n} > R)
\,.
}
Applying then Markov's inequality to the two last terms, one gets
\bea{
\Expectation\msp{1}j_n(w) 
\;\gtrsim\; 
&\Prob(\msp{1}\abs{X_n^\vartheta}_\T \leq w^2)
\,-\, \Prob(|L_n| > 1) 
\,-\, \Prob(|K_n| > 1)
\\
&-\frac{1}{R} \Expectation\Bigl[
\chi_{\Ball(0,w^2)}(X_n^\vartheta) \cdot  \Gamma_n^\vartheta
\Bigr]
\,-\, 
\frac{1}{R} \Expectation\Bigl[
\chi_{\Ball(0,w^2)}(X_n^\vartheta) \cdot  \nE^{M_n}
\Bigr]
\,. 
}
Proposition \ref{prp:Potential theory} and Lemma \ref{lmm:bound for K_n and L_n probabilities} allow then to conclude that $ \tE(j_n(w))\gtrsim w^2$ if $R$ is chosen large enough. This finishes the proof.

\subsection{Proof of the upper bound}
\label{ssec:Proof of the upper bound}

Let $ n \in \N$. Let $ c>0 $ to be fixed later. Starting from \eqref{CL-current and def of J_n}, one writes
\begin{equation}
\label{preuve upper}
\Expectation\msp{1}J^{\text{CL}}_n 
\;\sim 
 \int_0^{c/n} \Expectation\msp{1}j_n(w) \dif w 
 \,+  
 \int_{c/n}^{w_0} \Expectation\msp{1} j_n(w) \dif w 
 \,+
 \int_{w_0}^{\infty } \Expectation\msp{1}j_n(w)\dif w 
\;=:\; \mathcal{J}_1 + \mathcal{J}_2 + \mathcal{J}_3
\,,
\end{equation}
with $ j_n $ defined in \eqref{def of jn}. Using the crude bounds $ D^2_{n-1}(e_1), D^2_n(e_2), D^2_{n-1}(e_2) \ge 0 $ in the definition of $ j_n $, and applying then Corollary \ref{crl:Fundamental decompositions of D_1n and D_2n}, one obtains
\bels{final proof: jn upper}{
j_n(w) 
\;\lesssim\;  
\frac{1}{1 + w^{-2}D_n^2(e_1)} 
\;\lesssim\; 
h(\Gamma_n^\vartheta\, \sin \pi X_n^\vartheta) 
\quad\text{with}\quad
h(r) \;=\; \frac{1}{1\,+w^{-4}r^2}
\,.
}
Let us first bound $ J_1 $. Let $w\in \lbrack 0 , c/n\lbrack$. First, $\Gamma_{n}^\vartheta \gtrsim 1 $,  as can be checked from its definition \eqref{def of Gamma^x_n}. Next, if $ c $ is small enough, one has, by point (i) of Corollary \ref{crl:the three qualitative properties of X}, that
\[
wn \;\lesssim\; X_n \;\leq\; \frac{1}{2}wn \;\leq\; \frac{1}{2}
\,.
\]
Therefore one has $\sin^2\pi X_n^\vartheta \,\gtrsim w^2n^2$, and thus
\bels{preuve upper J1}{
\mathcal{J}_1 \;\lesssim\; \int_0^{c/n}\frac{\dif w}{1 \,+\, w^{-2} n^2 } 
\;\lesssim\; 
n^{-3}
\,.
}
%
Let us next bound $ \mathcal{J}_2$. Let $ w \in \lbrack c/n, w_0 \lbrack$, and $ m = \min \lbrace n, \lfloor w^{-2}\rfloor \rbrace$. One writes
\bels{final proof: decomposition jn}{
\Expectation\msp{1}j_n(w)
\;=\; 
\int_\R \int_\T \Expectation \big( j_n(w) | X_{n-m}^\vartheta = x, \Gamma_{n-m}^\vartheta = a \big) \, 
\Prob (X_{n-m}^\vartheta \in \dif x, \, \Gamma_{n-m}^\vartheta \in \dif  a)
\,.
}
To simplify notations, set $ \tE(\genarg|x,a) := \Expectation (\genarg | X_{n-m}^\vartheta = x, \Gamma_{n-m}^\vartheta = a) $.
If $x\in \T$ and $a \in\R$ are given, it follows from \eqref{final proof: jn upper} that
\bels{final proof:conditional jn}{
\Expectation(j_n(w)|x,a) 
\;\lesssim\; 
\Expectation 
\msp{1}
h(a\msp{1} \Gamma_m^x\sin \pi X_m^x)
\,, 
}
since, by the definition \eqref{def of Gamma^x_n}, one may write $ \Gamma_n^\vartheta = \prod_{l=1}^n g(X_{l-1}^\vartheta, B_l)  =  \Gamma_{n-m}^\vartheta \prod_{l=n-m+1}^n g(X_{l-1}^\vartheta, B_l)$, for some function $g$.
Because $h(r) \le 1$ and $h(r)\le w^{4}r^{-2}$ for every $r\in\R$, one has, for every event $A$, the bound 
\bels{final proof:decomposition hr}{
h(a\, \Gamma_m^x \sin \pi X_m^{x} )  
\;\leq\;  
1_A \,+\, 1_{A^c} \cdot  w^4 \cdot (a\, \Gamma_m^x \sin \pi X_m^{x})^{-2} \, .
}
So, taking $ 1_A = \chi_{\lbrack0,1\rbrack }(w^{-4}a^2\sin^2 \pi X_m^x)$, and using \eqref{final proof:decomposition hr} in \eqref{final proof:conditional jn} one obtains
\[
\Expectation(j_n(w)|x,a) 
\;\lesssim\; 
\Expectation\Bigl\{
\chi_{\lbrack 0,1\rbrack} (w^{-4}a^2\sin^2 \pi X_m^x) \,+ \, \chi_{\rbrack 1,\infty\lbrack} (w^{-4}a^2\sin^2 \pi X_m^x) \cdot w^{4} \cdot (a\, \Gamma_m^x \sin \pi X_m^x)^{-2}
\Big\}
\,.
\]
Therefore, Proposition \ref{prp:Potential theory} implies
\bea{
\Expectation(j_n(w)|x,a)
\;&\lesssim\;  
\frac{1}{w\sqrt{m}}\int_\T 
\bigl\{ 
\chi_{\lbrack 0,1\rbrack} (w^{-4}a\sin^2\pi y) \,+\,  \chi_{\rbrack 1,\infty\lbrack}(w^{-4}a\sin^2 \msp{-2}\pi y) \, w^{4}a^{-2} \sin^{-2} \msp{-3}\pi y 
\bigr\}
\dif y
\\
&\lesssim\; 
\frac{1}{w\sqrt{m}} \int_\T  \frac{\dif y}{1 \,+\, w^{-4}a^2\sin^2 \pi y}
\;\lesssim\; 
\frac{1}{w\sqrt{m}} \int_{-1/2}^{1/2} \frac{\dif y}{1 \,+\, (w^{-2}a \msp{1}y)^2} 
\\
&\leq\;  
\frac{w^2 a^{-1}}{w\sqrt m }\int_{-\infty }^{+\infty } \frac{\dif z}{1 + z^2} \;\lesssim \; 
\frac{w}{\sqrt m}\,a^{-1} 
\,,
}
where one has used the change of variables $z = w^{-2}ay$ to get the third line.
One now inserts this last bound in \eqref{final proof: decomposition jn}.
Applying Proposition \ref{prp:expectation of 1/Gamma_n decays exponentially}, one gets
\bea{
\Expectation\msp{1}j_n (w)   
\;&\lesssim\;
\int_\R \int_\T \frac{w}{\sqrt m} \,a^{-1} \, \Prob (X_{n-m}^\vartheta \in \dif x, \,\Gamma_{n-m}^\vartheta \in \dif  a)
\\
&=\;\frac{w}{\sqrt m }\Expectation(1/\Gamma_{n-m}^\vartheta) \lesssim \frac{w}{\sqrt m } \nE^{-\alpha w^2 (n-m) } 
\;\lesssim\;  
\max \!\left\{\!\frac{w}{\sqrt {n}}, w^2\right\} \nE^{-\alpha w^2 n }.
}
Therefore 
\bels{final proof: J2 upper}{
\mathcal{J}_2 \;\lesssim\; 
\frac{1}{\sqrt n} \int_0^{ n^{-1/2} } w\msp{1}\nE^{-\alpha w^2 n} \dif w 
\,+\, \int_{n^{-1/2}}^\infty w^2 \nE^{-\alpha w^2 n } \dif w 
\;\lesssim\; 
n^{-3/2}
\,.
}
It has already been shown by O'Connor \cite{O'Connor-75} that $ \mathcal{J}_3 \lesssim  \nE^{-C n^{1/2}}$. 
One thus finishes the proof by inserting this last estimate, together with \eqref{preuve upper J1} and \eqref{final proof: J2 upper} in \eqref{final proof: jn upper}.

\subsection{On other heat baths}
\label{ssec:Other heat baths}

Associate a heat bath to a function $ \mu : \R \to \C $ as described by Dhar \cite{Dhar_Spect_Dep-01}. One may then obtain, at least formally, a new heat bath by replacing $ \mu $ with a function $ \tilde{\mu} : \R \to \C $ defined by scaling $ \tilde{\mu}(w) \sim \mu(\mathrm{sgn}(w)\abs{w}^s) $, $ s > 0 $. 
In \cite{Dhar_Spect_Dep-01} Dhar argued based on numerics and a non-rigorous approximation that Casher-Lebowitz and Rubin-Greer bath functions $ \mu_{\mathrm{CL}}(w) \sim \cI w $ and $ \mu_{\mathrm{RG}}(w) \sim \nE^{-\cI \pi \vartheta(w)} $, with $ \vartheta(w) $ given in \eqref{def of average shift}, yield $ \tE\msp{1} J^{\wti{\mathrm{CL}}}_n \sim n^{-(1+s/2)} $ and $ \tE \msp{1}J^{\wti{\mathrm{RG}}}_n \sim n^{-(1+\abs{s-1})/2} $, respectively. 
The first of these statements can be proven rigorously by directly adapting the proof of Theorem \ref{thr:the scaling of the average current}. 
The second case, however, does not follow directly from the proof of $ \tE\msp{1}J^{\mathrm{RG}} \sim n^{-1/2} $, even though we believe it should not be too difficult to prove by using our results.

To see where the difficulties within this second case lie, as well as to further demonstrate our approach, let us sketch how $ \tE\msp{1}J^{\mathrm{RG}} \sim n^{-1/2} $, first proven by Verheggen \cite{Verheggen-1979}, can be obtained by using our representation of $ D_n(v) $.
Indeed, the choices $ \tilde{e}_1 := 2^{-1/2}(e_1 + e_2)$ and $ \tilde{e}_2 := 2^{-1/2}(e_1 - e_2) $ yield (Proposition \ref{prp:Fundamental decomposition of D_n(v)}) $ D_n(\tilde{e}_1) \sim \Gamma^{x_1}_n \sin \pi X^{x_1}_n $ and $ D_n(\tilde{e}_2) \sim w^{-1}\Gamma^{x_2}_n \sin \pi X^{x_2}_n $ with $ x_1 = 1/2 + \mathcal{O}(w) $ and $ x_2 = w/2 + \mathcal{O}(w^2) $, respectively. 
If one substitutes these in the expression for the current density $ j^{\mathrm{RG}}_n(w) $ of the Rubin-Greer model (the equation between 3.1 and 3.2 in \cite{Verheggen-1979}) one ends up with an estimate 
\bels{RG-current current density}{
(1+(\Gamma^{x_1}_n)^2 + (\Gamma^{x_2}_n)^2)^{-1} 
\;\lesssim\; 
j^{\mathrm{RG}}_n(w) 
\;\lesssim\;
(1+(\Gamma^{x_2}_n)^2)^{-1}\,,
\quad\text{for}\quad w \leq w_0
\,,
}
after making use of the basic properties of $ X $-processes (Corollary \ref{crl:the three qualitative properties of X}). This reveals that the Rubin-Greer model is special in the sense that the random phases $ X^{x_k}_n $ in the expressions $ D_n(\tilde{e}_k) \sim  \Gamma^{x_k}_n \sin \pi X^{x_k}_n $ do not have any direct role in the scaling behavior of the current. 
The reason why proving $ \tE\msp{1} J^{\wti{\mathrm{RG}}}_n \sim n^{-(1+\abs{s-1})/2} $, $ s \neq 1 $, is again more difficult is that the bounds analogous to \eqref{RG-current current density} become again explicitly depended on $ X^{x_k} $. 

Now continuing with the RG-model, based on \eqref{RG-current current density} one can prove $ \tE\msp{1}j^{\mathrm{RG}}_n(w) \sim \nE^{-C w^2 n} $  which then implies the scaling: $ \tE J^\mathrm{RG}_n = \int_\R \tE \msp{1}j^{\mathrm{RG}}_n(w) \dif w \sim n^{-1/2} $. 
Indeed, for the lower bound $ \tE\msp{1}j^{\mathrm{RG}}_n(w) \gtrsim  \nE^{-C w^2 n} $ one considers the typical behavior, which is easier to analyze than in the Casher-Lebowitz model since $ X $-processes are not present. 
The respective upper bound follows from Proposition \ref{prp:expectation of 1/Gamma_n decays exponentially}.

\appendix
\section{Appendix}

\subsection{Proof of Lemma \ref{lmm:functions fb and Phi}}
\label{assec:Proof of fb-lemma}

By using \eqref{Mobius maps preserve matrix multiplication} one gets 
\bea{
f_b  
\;\equiv\;  
g^{-1} \circ \mathcal{M}_{A} \circ g
\;=\; 
E \,\circ \mathcal{M}_{G^{-1}\!A G} \circ E^{-1}
\,,
}
where 
\bels{wtiLambda}{
G^{-1}A G 
\;=\;
\mat{(1 + \cI\delta) \nE^{\cI \pi \vartheta} & -\cI\delta\msp{1} \nE^{\cI \pi \vartheta} \\ 
\cI\delta\msp{1} \nE^{-\cI \pi \vartheta} & (1-\cI \delta) \nE^{-\cI \pi \vartheta} }
\,,
}
and 
\bels{def of delta}{
\delta 
\;=\; 
\frac{\pi^2 w^2 b}{2 \sin \pi \vartheta}
\;=\;
\frac{(\pi w/2)b}{\sqrt{1-(\pi w/2)^2}}
\;=\; 
(\pi w/2)\,b \,+\, \mathcal{O}(w^3b) 
\,.
}
Here the second equality follows from \eqref{def of average shift}. 

The map $ \mathcal{M}_{G^{-1} A G} $ describes the evolution $ \xi \mapsto \mathcal{M}_A(\xi) $ on the complex unit circle $ \partial D $:
\[ 
\mathcal{M}_{G^{-1} A G}(\nE^{\cI \phi}) \;=\; \nE^{\cI (\phi + 2\pi \vartheta)}\, \frac{1\,+\,\cI \delta\, (1-\nE^{-\cI \phi})}{1\,-\,\cI \delta\, (1-\nE^{\cI \phi})} 
\;=:\; \exp\bigl[\cI (\phi + 2\pi \vartheta + 2 \tilde{\Phi}(\phi,\delta))\bigr]
\,.
\]
Here the effect of noise $ \delta $ comes through   
\bels{delta zero-average random shift}{
\tilde{\Phi}(\phi, \delta) 
\;&=\;
\arg \bigl[ 1 + \cI \delta (1-\nE^{-\cI \phi})\bigr]
\;=\; \arctan\left[ \frac{1-\cos \phi}{1 - \delta \sin \phi}\, \delta \right]
\\
&=\; (1-\cos \phi)\, \delta 
\; + \; 
(1-\cos \phi) \sin \phi \,\delta^2  
\;+\; 
\mathcal{O}\bigl((1-\cos \phi)\delta^3\bigr) 
\,.
}
By substituting $ \phi = 2\pi x $ and using the middle expression of \eqref{def of delta} in place of $ \delta $ we obtain \eqref{Phi as arctan}. 

Let $ h(w,x,b) $ be a function so that $ w b\, h(w,x,b) \sin^2 \!\pi x  $ equals the argument of $ \arctan $ in \eqref{Phi as arctan}. It is easy to see that $ h $ is a smooth bounded function on $ [0,w_0] \times \T \times [b_-,b_+] $.
We may then write $ \Phi(x,b) \equiv \Phi(w;x,b) $ as
\bels{finding the product form for Phi}{
\Phi(w,x,b) 
\;=\; 
\frac{1}{\pi} w b\, h(w,x,b) \sin^2\!\pi x
\,+\, \frac{1}{6\pi} \arctan'''(s) \bigl[ w b\, h(w,x,b) \sin^2\!\pi x \bigr]^3
\,,
}
where the third derivative $ \arctan'''(s) $ of $ \arctan $ is bounded on $ 0 \leq s \leq w b \sin^2(\pi x) h(w,x,b) = \mathcal{O}(w) $.
By expanding $ h(w,x,b) = \pi + \mathcal{O}(w) $ similarly, and then substituting the result back into \eqref{finding the product form for Phi} one obtains \eqref{Phi in product form}.

To prove the formula \eqref{inverse of f_b in terms of Phi} for $ f^{-1}_b $ we note that $ \nE^{\cI 2 \pi f_b^{-1}(y)} = \mathcal{M}_{\wti{\Lambda}}^{-1}(\nE^{\cI 2 \pi y}) = \mathcal{M}_{\wti{\Lambda}^{-1}}(\nE^{\cI 2 \pi y}) $ where $ \wti{\Lambda} $ is the matrix in \eqref{wtiLambda}. After replacing $ \wti{\Lambda} $ by its inverse, the proof proceeds just like before. 
The identity involving $ \Phi(x,-b) $ follows by expressing $ f_b $ and $ f_b^{-1} $ in terms of $ \Phi $ in $ x = f_b^{-1}(f_b(x)) $. 
%

\subsection{Proof of Lemma \ref{lmm:Freedman's and Azuma's martingale exponent bounds}}
\label{assec:Proofs of Freedman's and Azuma's bounds}


Both proofs are rather directly adapted from Freedman's paper \cite{Freedman-75}.
We start with Freedman's bound.
To this end define a function $ g: \R \to \R $: $ g(0)=1/2 $, $ g(t) := (\nE^t-1-t)/t^2 $ for $ t \neq 0 $. Let $ t,y \in \R $ so that $ \abs{y} \leq 1 $. By definition we have then
\[
\nE^{ty} \;=\; 1 \,+\, ty \,+\, (ty)^2g(ty)
\,.
\] 
It is not too difficult to see that $ g $ is an increasing function. Therefore, $ g(ty) \leq g(t) $ above, and
\bels{approximation of exponential}{
\nE^{ty} 
\;\leq\; 
1 \,+\, ty \,+\, y^2 t^2g(t) 
\;=\; 
1 \,+\,ty\,+\,y^2(\nE^t-1-t) 
\;\equiv\; 
1\,+\,ty\,+\,y^2 \kappa_1(t)
\,.
}
Suppose $ Y $ is a random variable such that $ \abs{Y} \leq 1 $ and $ \tE(Y) = 0 $. Setting $ y = Y $ in \eqref{approximation of exponential} and taking expectation yields 
\bels{main relation behind in Freedman}{
\tE\,\nE^{tY} 
\;&\leq\; 
\tE\bigl(1\,+\,tY\,+\,\kappa_1(t) Y^2 \bigr)
\;=\; 
1 \,+\,\kappa_1(t)\tE(Y^2)
\;\leq\; 
\nE^{\kappa_1(t)\tE(Y^2)}
\,.
} 
Now, set $ Y_i := (M_i-M_{i-1})/m $, so that $ \abs{Y_i} \leq 1 $ and $ \tE(Y_i|\mathcal{F}_{i-1}) = 0 $. By using $ \kappa_m(t)= m^{-2}k_1(tm) $ to write $ \kappa_m(t) (M_i-M_{i-1})^2 = \kappa_1(mt) Y_i^2 $, the estimate \eqref{main relation behind in Freedman} implies that for any $ t \in \R $:  
\bels{the main bound of Freedman}{
\tE\Bigl(\nE^{t(M_i-M_{i-1})\,-\,\kappa_m(t)\tE[(M_i-M_{i-1})^2|\mathcal{F}_{i-1}]}\Big|\mathcal{F}_{i-1}\Bigr) 
\;=\;
\tE\Bigl(\nE^{tm Y_i\,-\,\kappa_1(tm)\tE[Y_i^2|\mathcal{F}_{i-1}]}\Big|\mathcal{F}_{i-1}\Bigr)
\;\leq\; 1
\,.
}
Recall the definition \eqref{def of V_n} of $ V_n $ and the pointwise bound $ V_n \leq v_n $. Apply these to get the first two lines below. Then use \eqref{the main bound of Freedman} iteratively to get Freedman's bound:
\bea{
\tE\,\nE^{tM_n} \;
&\leq\;
\nE^{\kappa_m(t)v_n}\tE\,
\nE^{tM_n-\frac{1}{2}\kappa_m(t)V_n}
\\
&=\;
\nE^{\kappa_m(t)v_n}\tE\Bigl\{ \nE^{tM_{n-1}-\frac{1}{2}\kappa_m(t)V_{n-1}}
\tE\Bigl(\nE^{t(M_n-M_{n-1})\,-\,\kappa_m(t)\tE[(M_n-M_{n-1})^2|\mathcal{F}_{i-1}]}\Big|\mathcal{F}_{i-1}\Bigr)
\Bigr\}
\\
&\leq\; \nE^{\kappa_m(t)v_n} \tE\, \nE^{tM_{n-1}-\frac{1}{2}\kappa_m(t)V_{n-1}}
\;\leq\; \cdots \;\leq\; \nE^{\kappa_m(t)v_n}
\,.
}
The bound \eqref{def of kappa_m} comes from the power expansion $ k_m(t) = (1/2) t^2 + k_m'''(s)t^3 =  (1/2)t^2 + (m/6)\msp{1} \nE^{ms}t^3 $, with $ s \in [0,t] $, by taking $ s = \abs{t} $. 

The proof of Azuma's bound proceeds in a very similar way: First, one uses the convexity of the exponent function to get a bound
\bea{ 
\nE^{ty} 
\;&=\; 
\nE^{\frac{1+y}{2}t \,+ \frac{1-y}{2}(-t)}
\;\leq\;
\frac{1+y}{2}\nE^{t} \,+\, \frac{1-y}{2} \nE^{-t}
\;=\; \cosh t \,+\, y \sinh t
\;\leq\; 
\nE^{t^2/2} +\, y \sinh t
\,,
}
for every $ t, y \in \R $ with $ \abs{y} \leq 1 $.
Using this instead of \eqref{approximation of exponential} in the first inequality of \eqref{main relation behind in Freedman} yields the bound $ \tE\, \nE^{tY} \leq \nE^{\frac{1}{2}t^2} $, and consequently $ \tE\bigl(\nE^{t(M_i-M_{i-1})}\big|\mathcal{F}_{i-1}\bigr) = \tE\bigl(\nE^{(tm)Y_i}\big|\mathcal{F}_{i-1}\bigr) \leq \nE^{(tm)^2/2} $. 
Iterating this finishes the proof:
\[
\tE\msp{1}\nE^{tM_n} \;=\; 
\tE\bigl\{ \nE^{tM_{n-1}} \tE(\nE^{t(M_i-M_{i-1})}|\mathcal{F}_{n-1})\bigr\} 
\;=\; 
\nE^{t^2m^2/2} \tE( \nE^{t M_{n-1} }) 
\;\leq\; 
\cdots 
\;\leq\; \nE^{t^2 m^2n/2}
\,.
\]

\subsection{Proof of Lemma \ref{lmm:approximate kernels}}
\label{assec:proof for approximate kernels}

Let us start with some conventions and definitions:
For $ k \in \N_0$, $ y \in \T$ we define:
\[
y_k : = y - kw, \quad
\alpha_k := \phi (y_k), \quad 
\gamma_k := h(y_k)
\,.
\]
For $ \epsilon > 0 $ and $ n_0 \in \N$, one defines 
\[
H(\epsilon,n_0) \;:=\; \setb{ (y,n)\in\mathbb T \times\mathbb N : |y|_\mathbb T \ge \epsilon, n\ge n_0,\, w^2n \ge \epsilon 
} 
\,.
\]
For $ u \in \Lspace^1(\T) $ and $ \xi \in \Z$, one defines 
\[
\hat{u}(\xi) = \int_\mathbb T u (x) \nE^{-\cI 2\pi \xi x }\dif x
\,.
\]

The operators $T_{y_k}$ $(k\in\mathbb N)$ are diagonal in Fourier space: 
for every $ \xi \in \Z $, one has
\bels{diagonalization Tk}{
\widehat{(T_{y_k}\!u)}(\xi) \;=\; \nE^{\cI 2 \pi w \xi}\lambda_k (w\xi)  \cdot \hat{u}(\xi)
\,,
}
where $\lambda_k$ is a function on $\mathbb R$ defined by
\bels{lambda}{
\lambda_k(z) \;:&= 
\int \nE^{\cI 2\pi zw^{-1}(\vartheta - w + \Phi(y_k,b)) } (1 + w \gamma_k b)\tau (b)\dif  b 
\\
&= 
\int \nE^{\cI 2 \pi z(\alpha_kb + \mathcal O(w))}(1 + w \gamma_k b)\tau (b)\dif  b .
}
Let $y\in\mathbb T$, let $u\in\mathrm L^1_{\Ball(y,w^2) }(\mathbb T;\mathbb R_+)$, and let $v\in \mathrm L^1_{\Ball(0,w^2) }(\mathbb T;\mathbb R_+)$ be such that 
\bels{u and v}{
v(x)=u(x+y).
}
One writes
\bels{qn fourier}{
S_{y,n} u (x) \;=\; T_{y_n} \cdots T_{y_1} v(x-y) \;=\; \sum_{\xi\in\Z }\nE^{\cI 2\pi\xi (x+nw -y) } \Lambda_n(\xi) \hat{v}(\xi).
}
where $\Lambda_n$ is a function on $\mathbb R$ defined by
\bels{Pn}{
\Lambda_n (\xi) := \prod_{j=1}^{n} \lambda_j (\xi w) \quad (n\ge 1).
}
But, 
if $(y,n)\in H(\epsilon, 8)$ for some $\epsilon>0$,
the right hand side of \eqref{qn fourier} represents actually a $ \Cont^2$-function. 
This follows directly from \eqref{Pn grand} with $l=0$ in Lemma \ref{lmm:lambda} below, and the fact that $|\hat{v}(\xi) |\le \norm{v}_1$ for every $\xi\in\Z$.

\Lemma{lambda}{
Let $\epsilon >0$.
There exist $K,K',\epsilon'>0$ such that, for every $(y,n)\in H(\epsilon, 1)$, 
and for every $\xi \in\mathbb R$ satisfying $|\xi w|\le \epsilon'$,
one has
\begin{align}
&\nE^{-Knw^2\xi^2 } \le |\Lambda_n(\xi)| \le \nE^{-K'nw^2\xi^2 }, \label{Pn gauche droite}\\
&|\Lambda'_n(\xi) | \le K nw^2 (1+|\xi|) \nE^{-K'nw^2\xi^2 },\label{Pn derivee}\\
&|\Lambda''_n(\xi) | \le  Knw^2 (1 + nw^2 + nw^2 \xi^2)\nE^{-K'nw^2\xi^2 }, \label{Pn 2 derivee}\\
&|\arg(\Lambda_n(\xi))| \le K nw^2 (|\xi | + w |\xi |^3).\label{Pn phase}
\end{align}
For every $\epsilon'>0$, there exist $K,K' >0$ such that, for every $(y,n)\in H(\epsilon,1)$, 
and for every $\xi \in\Z$ satisfying $|\xi w | >\epsilon'$, one has
\bels{Pn grand}{
\abs{\partial_\xi^l \Lambda_n(\xi)} 
\;\leq\; 
\frac{K(wn)^l}{(1+ K' |\xi w |)^{n/2}}, \quad l=0,1,2.
}
}

\begin{Proof}
The constants introduced in this proof may depend on $\epsilon$.
For the whole proof, one sets $z = \xi w$.
Before starting, let us make two observations. 
First, one has $|\alpha_k | \lesssim 1$ and $|\gamma_k| \lesssim 1$ for every $k\in\mathbb N_0$.
Secondly, for every $(y,n)\in H(\epsilon,1)$, there exists an integer $m\ge n/2$ independent of $y$, and a subsequence
\bels{jk}{
\{k_j\} \;\equiv\;\sett{k_j : 1\le j\le m} \;\subset\; \sett{1,2,\dots , n}
}
such that $|\alpha_{k_j}|\gtrsim 1$.

Let us first prove the formulas (\ref{Pn gauche droite}) up to (\ref{Pn phase}).
One takes $(y,n)\in H(\epsilon,1)$.
A Taylor expansion in \eqref{lambda}, taking into account that $\int \tau (b)\dif  b = 1$ and $\int b\tau(b)\dif  b = \mathsf E(B)= 0$, gives
\[
\lambda_k(z) \;=\; 1 + \cI \mathcal O(w|z|) \,-\, \frac{(2\pi)^2}{2}z^2\alpha_k^2 
\tE(B^2) \,+\, \mathcal{O}(wz^2) \,+\, \cI \mathcal{O}(\abs{z}^3) \,+\, \mathcal O(z^4)\,, 
\]
as $z\to 0$.
Therefore, one has
\begin{align}
\abs{\lambda_k(z)} \;&=\; \nE^{-\frac{(2\pi)^2}{2}z^2 \alpha_k^2  \tE(B^2) \,+\, \mathcal{O}(z^2 w + \abs{z}^3)}
\,, 
\label{lambda-loc1}
\\
|\arg(\lambda_k(z)) | \;&=\; \mathcal O(|z|w + |z|^3)\label{lambda-loc4}
\,,
\end{align}
as $z\to 0$. Similarly, a Taylor expansion in \eqref{lambda} gives
\begin{align}
|\partial_z \lambda_k(z) | \;&=\;  \mathcal O(w + |z|)
\,, 
\label{lambda-loc2}
\\
|\partial_z^2 \lambda_k(z) | \;&=\; \mathcal O(1)
\,,
\label{lambda-loc3}
\end{align}
as $z\to 0$. 

First, by \eqref{lambda-loc1} with $z=\xi w$,  and by the definition \eqref{Pn} of $\Lambda_n$, one obtains
\[
\abs{\Lambda_n(\xi)} = 
\exp 
\biggl[ -\frac{1}{2}(\xi w)^2 \tE(B^2)\sum_{k=1}^{n}\alpha_k^2 + \mathcal O\big( n (\xi w )^2(|\xi w | + w)\big) 
\biggr]
\quad \mathrm {as} \quad \xi w \to 0
\,.
\]
This shows \eqref{Pn gauche droite}, taking into account the two observations at the beginning of this proof. 
Next, with $z=\xi w$, one has 
\begin{align}
\label{deriv 1}
\partial_\xi \Lambda_n(\xi) 
\;&=\; 
w \sum_{j=1}^{n} \partial_z \lambda_j(z) \prod_{\stackrel{1\le k\le n}{k\ne j} } \lambda_k(z)\,,
\\
\label{deriv 2}
\partial_\xi^2 \Lambda_n(\xi) 
\;&=\; 
w^2\sum_{j=1}^{n} \Big( \partial_z^2\lambda_j(z) \prod_{\stackrel{1\le k\le n}{k\ne j} } \lambda_k(z) + \partial_z \lambda_j(z)\sum_{\stackrel{1\le k\le n}{k\ne j} }\partial_z \lambda_k(z)\prod_{\stackrel{1\le l\le n}{l\ne j,k} }\lambda_l(z)\Big)
\,. 
\end{align}
One then obtains \eqref{Pn derivee} and \eqref{Pn 2 derivee}, by using these last formulas together with \eqref{lambda-loc2}, \eqref{lambda-loc3}, and the fact that $|\lambda_k (z) |\le 1$ for every $k \in \N$ and every $ z \in \R $,  which follows from  the definition \eqref{lambda}. Finally, \eqref{Pn phase} directly follows from \eqref{lambda-loc4}.

Let us now show \eqref{Pn grand}.
Let $\epsilon' >0$, and let $(y,n)\in H(\epsilon, 1)$.
The constants introduced below may depend on $\epsilon'$.
One proceeds in two steps. 

First, one shows \eqref{Pn grand} for $|z|=|\xi w|\in\lbrack \epsilon', 1/\epsilon'\lbrack $.
It is actually enough to show that 
\bels{appendix: lemm interm 1}{
|\lambda_k(z) |\le 1 - \epsilon_1 
}
for some $\epsilon_1 >0$ and for every $k \in \{k_j\} $, with $\{k_j\}$ as defined in \eqref{jk}.
Indeed, from the definition \eqref{lambda}, one has $|\lambda_k(z)|\le 1$ and 
$|\partial^l_z \lambda_k(z)| \lesssim 1 $ for $l=1,2$, for every $k \in \N $ and every $ z \in \R $. So, inserting \eqref{appendix: lemm interm 1} in \eqref{Pn}, \eqref{deriv 1} or \eqref{deriv 2}, respectively for $l=0$, $l=1$ or $l=2$, will imply
\[
\abs{\partial_\xi^l \Lambda_n(\xi)} 
\;\lesssim\; 
(wn)^l (1 - \epsilon_1)^{\frac{n}{2} - 2}
\,, 
\]
which is equivalent to \eqref{Pn grand} when $\epsilon' \le  |\xi w| < 1/\epsilon'$.

So let us show \eqref{appendix: lemm interm 1}.
By continuity of $\tau$, one finds an interval $J$ on which $\tau \ge \epsilon_2$ for some $\epsilon_2 > 0$.
One has 
\[
\int_J   \nE^{\cI z(\alpha_kb + \mathcal{O}(w) )}(1 + w \gamma_k b)\tau (b)\dif  b = \int_J \nE^{\cI z \alpha_k b } \tau (b)\dif  b + \mathcal O(w)
\,,
\]
and, for some $\epsilon_3 >0$,
\[
\abs{\int_J \nE^{\cI z \alpha_k b } \tau (b) \dif b} 
\leq 
(1 - \epsilon_3)\int_J \tau (b) \dif b
\,.
\]
Therefore
\[
\abs{\lambda_k (z)} 
\leq 
(1-\epsilon_3)\int_J \tau (b) \dif b + \int_{J^\cmpl}\tau (b)\dif  b + \mathcal O(w) =
1 - \epsilon_3 \int_J \tau (b) \dif b + \mathcal O (w) \le 1 - \epsilon_2 \epsilon_3 \mathrm{Leb}(J) + \mathcal O(w)
\,.
\]
One thus may take $\epsilon_1  = \frac{1}{2}\epsilon_2\epsilon_3 \mathrm{Leb}(J)$.

Next, one shows \eqref{Pn grand} for $|z|=|\xi w| \ge 1/\epsilon'$.
Here, it is enough to show that, for some $C>0$, one has
\bels{appendix: interm lemm a}{
\abs{\partial_z^l \lambda_k(z)} \,\leq\, C/\abs{z}
\,,
}
for $l=0,1,2$, and for every $k\in \{k_j\} $. Indeed, if $ \epsilon' $ has been taken small enough, one finds some $C'>0$ such that $C/|z|\le 1/(1 + C'|z|)$, when $|z|\ge 1/\epsilon' $. 
So, inserting now \eqref{appendix: interm lemm a} in \eqref{Pn}, \eqref{deriv 1} or \eqref{deriv 2}, respectively for $l=0$, $l=1$ or $l=2$, one will obtain \eqref{Pn grand} for  $| \xi w| \ge 1/\epsilon'$.

So let us show \eqref{appendix: interm lemm a}. It follows  from \eqref{lambda} that $\partial_z^l \lambda_k(z)$ can be written under the form $ \partial_z^l \lambda_k (z) = \int \nE^{\cI  z \mu(b)}\rho_l(b)\dif b $. An integration by parts gives
\[
\partial_z^l \lambda_k(z) 
\;=\; 
\frac{1}{z} \frac{\nE^{\cI z \mu (b) \rho (b)}}{\cI \partial_b \mu (b) } \Big|^{b_+}_{b_-}
-
\frac{1}{z} \int_{b_-}^{b_+} \nE^{\cI z\mu (b) } \partial_b \Big( \frac{\rho(b) }{\cI \partial_b \mu (b) }\Big) 
\dif b
\,. 
\]
Here, one has $\rho_l \in \Cont^1(\lbrack b_-, b_+\rbrack)$, since $\tau \in \Cont^1(\lbrack b_-, b_+\rbrack)$, 
and $|\partial_b^j \rho (b)| \lesssim 1 $ for $j=0,1$.
Moreover, one checks form the definition \eqref{def of Phi} of $\Phi$ that $|\partial_b \mu(b)| \gtrsim 1$ and $|\partial_b^2 \mu (b)| \lesssim 1$. This finishes the proof.
\end{Proof}

One now let $(y,n)\in H(\epsilon ,8)$. The constants introduced below  depend on $y$ only through $\epsilon$.
\begin{Proof}[Proof of \eqref{approximate kernels 1}]

By \eqref{qn fourier}, \eqref{Pn gauche droite} and \eqref{Pn grand} (with $l=0$),  
there exists $\epsilon' >0$ such that,
\bea{
|\partial_x^l S_{y,n}u(x) | \; 
&\leq\;  
(2\pi)^2 \sum_{\xi\in\Z } |\xi|^l |\Lambda_n(\xi)| \, \norm{u}_1
\\
&\lesssim\;  
\norm{u}_1  \sum_{\xi:|\xi w |\le \epsilon' }|\xi |^l \nE^{-C n (\xi w)^2}  +  \norm{u}_1 \sum_{\xi:|\xi w |> \epsilon' } \frac{|\xi |^l}{(1 + C | \xi w |)^{\frac{n}{2}} } 
\\
&\lesssim\;   
\frac{\norm{u}_1}{w^{l+1}}
\int_0^\infty y^l \nE^{-Cny^2} \dif  y \,+\,  \frac{\norm{u}_1}{w^{l+1}} \int_{\epsilon' }^\infty \frac{y^l\dif  y }{(1+C y)^{\frac{n}{2} }} 
\\
&=:\;  \frac{\norm{u}_1}{w^{l+1}} (I_1 +I_2)
\,. 
}
But one has  $ I_1 \lesssim n^{-(l+1)/2} $, and
\bels{appendix-loc-2}{
I_2 
\;\leq\; 
\frac{1}{C^l}\int_{\epsilon' }^\infty \frac{\dif  y }{(1 + Cy)^{\frac{n}{2}-l }} 
\;\leq\; 
\frac{1}{C^{l+1}(\frac{n}{2}-l-1) (1 + C\epsilon')^{\frac{n}{2}-l-1 }} 
\;\lesssim\; 
\nE^{-C(\epsilon')n}
\,.
}
This finishes the proof.
\end{Proof}

\begin{Proof}[Proof of \eqref{approximate kernels 2}]
We will only consider the case $k=2$ ; the case $k=1$ can be handled similarly, and turns out to be easier.
To simplify the notations, one writes 
\[
A \;:=\; \sin^2 \pi (x+nw-y) \cdot \partial_x^2 S_{y,n}u(x)
\,.
\]
We recall that the function $ v $ defined in \eqref{u and v} satisfies $v(x)= u (x+y)$. One has $\sin^2 z = \frac{1}{4}(2 - \nE^{\cI 2z } - \nE^{-\cI 2z})$, and thus, by \eqref{qn fourier}, one has
\bels{appendix-loc-01}{
A \;
&=\; -\pi^2 \Bigl\{2 - \nE^{\cI 2\pi(x+nw - y)} - \nE^{-\cI 2\pi (x+nw-y)} \Bigr\}
\sum_{\xi \in\Z } \xi^2 \nE^{\cI 2\pi\xi (x+wn-y)}
\Lambda_n(\xi) \hat{v}(\xi) 
\\
&=\; -\pi^2 \sum_{\xi \in \Z } \nE^{\cI 2\pi \xi\msp{1} (x+wn - y)} 
\Bigl\{ 2 \msp{1}\xi^2\Lambda_n(\xi)\hat{v}(\xi) \,-\, (\xi-1)^2\Lambda_n(\xi-1)\hat v(\xi-1)
\\
&\msp{308}
-\, (\xi+1)^2\Lambda_n(\xi +1)\hat v(\xi +1) \Bigr\}
\,.
}
Since 
\[
\abs{\hat{v}(\xi) - \hat{v}(\xi-1)} 
\;\leq\; 
\int_{\Ball(0,w^2) } |v(x)| \, |1 - \nE^{\cI 2\pi x } | \dif x 
\;\lesssim\; 
w^2 \norm{u}_1
\,,
\]
for every $\xi\in \Z$, one has, for every $\epsilon' >0 $,  
\bels{appendix: eq 2 A}{
\abs{A}\; 
&\lesssim\;
\norm{u}_1 \sum_{\xi \in \Z}
\Big| 2 \xi^2\Lambda_n(\xi) - (\xi-1)^2\Lambda_n(\xi-1)
 - (\xi+1)^2\Lambda_n(\xi +1) \Big| \,+\, (\xi  w)^2 |\Lambda_n(\xi) |
 \\
&\lesssim \, \norm{u}_1  \sum_{\xi\in\Z}
\Bigl\{ \xi^2\big|  2 \Lambda_n(\xi) -  \Lambda_n(\xi-1) - \Lambda_n(\xi+1) \big|    
\\
&\msp{100}+\, |\xi| \cdot \big| \Lambda_n(\xi-1) - \Lambda_n(\xi+1)\big| \,+\, (1 + (\xi  w)^2) |\Lambda_n(\xi)| \Bigr\}
\\
&\lesssim\; \norm{u}_1  
\sum_{\xi\in \Z } 
\Big\{ 
\xi^2 |\Lambda_n''(\xi_1(\xi))| 
\,+\, 
|\xi | \cdot|\Lambda_n'(\xi_2(\xi))| 
\,+\, 
(1 + (\xi w)^2) |\Lambda_n(\xi)| 
\Bigr\}  
\\
&=\;  \norm{u}_1  \sum_{\xi :|\xi w |\le \epsilon' } (\,\cdots) \;+ \sum_{\xi : |\xi w |>\epsilon' } (\,\cdots) 
\;=:\; 
\norm{u}_1 ( I_1 + I_2)
\,. 
}
The numbers $\xi_1 (\xi)$ and $\xi_2 (\xi)$ in \eqref{appendix: eq 2 A} are obtained by a Taylor expansion and satisfy $|\xi_1(\xi) - \xi| \le 2$ and $|\xi_2(\xi) - \xi|\le 2$.

If $\epsilon'$ is taken small enough, then, by \eqref{Pn gauche droite},  \eqref{Pn derivee} and \eqref{Pn 2 derivee}, and because $nw^2\le 1$ by hypothesis, one has
\bels{lambda loc f1}{
I_1 
\;\lesssim\;  
\int_0^\infty \bigl\{ 1 \,+\, (\xi w\sqrt n)^2 \,+\, (\xi w\sqrt n)^4\bigr\} 
\,\nE^{-C (\sqrt n w \xi)^2} \dif \xi 
\;\lesssim\; 
\frac{1}{w\sqrt{n}}
\,.
}
By \eqref{Pn grand}, one gets as for \eqref{appendix-loc-2}, 
\bels{lambda loc f2}{
I_2 
\;\lesssim\; 
\sum_{|\xi| w \ge \epsilon' } \frac{ (\xi w n)^2 + \xi w n + 1  }{(1 + C\xi w)^{\frac{n}{2} } }
\;\lesssim\;  
\frac{1}{w} \int_{\epsilon' }^\infty \frac{\big( (y n)^2 + y n + 1 \big)\dif  y  }{(1 + Cy)^{\frac{n}{2} } }
\;\lesssim\; \frac{1}{w} \nE^{-C'n}
\,.
}
Inserting \eqref{lambda loc f1} and \eqref{lambda loc f2} in \eqref{appendix: eq 2 A} gives the result. 
\end{Proof}

\begin{Proof}[Proof of \eqref{approximate kernels 3}]
Let $\epsilon >0$ be as small as we want. 
One takes $x,y\in\mathbb T$ such that $|x+nw - y|_\mathbb T \le 10 \epsilon$.
The constants introduced below do not depend on $\epsilon$.
We recall that the function $v$ defined in \eqref{u and v} satisfies $v(x)= u (x+y)$.
Starting from \eqref{qn fourier}, one obtains
\bels{lambda loc A4}{
S_{y,n}u(x)  \;\ge \sum_{\xi : |\xi |\le \epsilon^{-2/3} } \nE^{2\cI \pi\xi (x+nw -y) } \Lambda_n(\xi) \hat{v}(\xi) \,-\! \sum_{\xi  : |\xi |> \epsilon^{-2/3} } |\Lambda_n(\xi)| \, \norm{u}_1
\,.
}

On the one hand, mimicking the proof of  \eqref{approximate kernels 1} with $l=0$, 
and taking the hypothesis $nw^2 \ge \epsilon$ into account, 
one finds, for some $\epsilon'>0$, 
\bels{lambda loc A3}{
\sum_{\xi  : |\xi |> \epsilon^{-2/3} } |\Lambda_n(\xi)|\;
&\lesssim\; 
\frac{1}{w}\int_{\epsilon^{-2/3}w }^{\epsilon' } \nE^{-Cny^2}\dif  y  + \frac{1}{w}\int_{\epsilon' }^\infty \frac{\dif y}{(1 + Cy)^{\frac{n}{2} }}
\\
&\lesssim\; 
\frac{1}{\sqrt \epsilon }\int_{\epsilon^{-1/6} }^\infty \nE^{-Cz^2}\dif z + \frac{1}{\sqrt \epsilon } \nE^{- C'(\epsilon')n} 
\; \lesssim\; 
\nE^{-C''\epsilon^{-1/3}} 
\,,  
}
where, to get rid of the term $\frac{1}{\sqrt \epsilon } \nE^{- C'(\epsilon')n}$, 
one has used the hypothesis $nw^2 \ge \epsilon$, which implies $n\ge \epsilon^{-1/3}$ when $w$ is small enough.

On the other hand, $\Lambda_n(-\xi) = \Lambda_n^*(\xi)$ by \eqref{Pn}, $\hat{v}(-\xi) = \hat{v}^*(\xi)$ since $u$ is real, and $\hat{v}(0)=\norm{u}_1$ since $u\ge 0$. Therefore
\bels{lambda loc A2}{
&\sum_{\xi : |\xi |\le \epsilon^{-2/3}  } \nE^{2\cI \pi\xi (x+nw -y) } \Lambda_n(\xi) \hat{v}(\xi)
\\
&\msp{10}=\; \norm{u}_1\,+\, 2\msp{-10}\sum_{1\le \xi \le \epsilon^{-2/3}  }|\Lambda_n(\xi)|\, |\hat{v}(\xi) |\cos \arg \bigl[ \nE^{2\cI \pi\xi (x+nw -y)} \Lambda_n(\xi)\hat{v}(\xi)\bigr]
\,.
}
Since $v\in\mathrm L^1_{\Ball(0,w^2) }(\T)$, one has $\arg (\hat{v}(\xi)) \lesssim |\xi |w^2$ for every $\xi\in\Z$.
So, by \eqref{Pn phase} and the hypothesis $|x+nw - y|_\T \le 10\epsilon$, one obtains
\bels{lambda loc A1}{
\big| \arg\big(\nE^{2\cI \pi\xi (x+nw -y)} \Lambda_n(\xi) \hat{v}(\xi)\big) \big| \;\lesssim\; 
\epsilon \xi  
\;\lesssim\;  \epsilon^{1/3}
\,,
}
when $1\le \xi \le \epsilon^{-2/3}$. But, if $1\le \xi \le \epsilon^{-2/3}$, one has
\bels{lambda loc Bz}{
|\hat v (\xi) | \ge \int v (x)\dif  x - \int_{\Ball(0,w^2) } |v(x)| \, |\nE^{-\cI 2\pi \xi x } -1| \dif  x  \ge (1 - C\xi w^2)\norm{u}_1 \ge \frac{1}{2}\norm{u}_1
\,,
}
and, by \eqref{Pn gauche droite}, one has $|\Lambda_n (\xi) | \ge \nE^{-Cnw^2 \xi^2}\ge \nE^{-C \epsilon \xi^2 }$, since $nw^2 \ge \epsilon$.
Therefore, using this last estimate, \eqref{lambda loc A1} and \eqref{lambda loc Bz} in \eqref{lambda loc A2} gives
\bels{lambda loc A2 b}{
\sum_{\xi : |\xi |\le \epsilon^{-2/3}  } \nE^{2\cI \pi\xi (x+nw -y) } \Lambda_n(\xi) \hat{v}(\xi) \gtrsim  \sum_{|\xi |\le  \epsilon^{-2/3}} \nE^{-C'\epsilon\xi^2 } \norm{v}_1
\,, 
}
if $\epsilon$ is small enough.

Therefore, inserting \eqref{lambda loc A3} and \eqref{lambda loc A2 b} in \eqref{lambda loc A4}, one gets
\[
S_{y,n}u(x) \;\ge\; \norm{u}_1 
\Bigl( 
C_1  \sum_{|\xi |\le  \epsilon^{-2/3}} \nE^{-C'\epsilon\xi^2 } - C_2 \, \nE^{-C''\epsilon^{-1/3}}\Bigr)
\,,
\]
and this tends to $\infty$ as $\epsilon \to 0$. 
\end{Proof}

\vspace{0.5cm}
\noindent {\bf Acknowledgments: }
A. Kupiainen deserves a special acknowledgement for introducing this problem to us and never sparing his time for enlightening comments and ideas.
We are grateful to J. Bricmont for helpful discussions and valuable feedback. We benefited from various illuminating discussions with M. Jara, J. Lukkarinen, M. Pakkanen, W. de Roeck, L. Saloff-Coste, A. Raugi and C. Liverani.
We both thank the Academy of Finland for Financial support. Additionally, O. Ajanki thanks European Research Council and F. Huveneers thanks the Belgian Interuniversity Attraction Poles Program for additional financial support.

\bibliography{1ddhc}

\begin{thebibliography}{10}

\bibitem{Anderson-58}
{\sc Anderson, P.~W.}
\newblock Absence of diffusion in certain random lattices.
\newblock {\em Phys. Rev. 109}, 5 (Mar 1958), 1492--1505.

\bibitem{RDS-Arnold-1998}
{\sc Arnold, L.}
\newblock {\em Random Dynamical Systems}.
\newblock Springer, Berlin, 1998.

\bibitem{Azuma-1967}
{\sc Azuma, K.}
\newblock Weighted sums of certain dependent random variables.
\newblock {\em Tohoku Mathematical Journal 19}, 3 (1967), 357--367.

\bibitem{Challange2000}
{\sc Bonetto, F., Lebowitz, J.~L., and Rey-Bellet, L.}
\newblock Fourier's law: a challenge to theorists.
\newblock In {\em Mathematical physics 2000}. Imp. Coll. Press, London, 2000,
  pp.~128--150.

\bibitem{Casher-Lebowitz-71}
{\sc Casher, A., and Lebowitz, J.~L.}
\newblock Heat flow in regular and disordered harmonic chains.
\newblock {\em Journal of Mathematical Physics 12}, 8 (1971), 1701--1711.

\bibitem{Coulhon-1990}
{\sc Coulhon, T., and Saloff-Coste, L.}
\newblock Puissances d'un op{\'e}rateur r{\'e}gularisant.
\newblock {\em Ann. Inst. H. Poincar{\'e}, Section B 26}, 3 (1990), 419--436.

\bibitem{Coulhon-1993}
{\sc Coulhon, T., and Saloff-Coste, L.}
\newblock Minoration pour les cha{\^i}nes de markov unidimensionnelles.
\newblock {\em Probability Theory and Related Fields 97\/} (1993), 423--431.

\bibitem{Dhar_Spect_Dep-01}
{\sc Dhar, A.}
\newblock Heat conduction in the disordered harmonic chain revisited.
\newblock {\em Phys. Rev. Lett. 86}, 26 (Jun 2001), 5882--5885.

\bibitem{Dhar-review-2008}
{\sc Dhar, A.}
\newblock Heat transport in low-dimensional systems.
\newblock {\em Adv. in Phys. 57}, 5 (Sep 2008), 457--537.

\bibitem{Escaurazia-2000}
{\sc Escaurazia, L.}
\newblock Bounds for the fundamental solution of elliptic and parabolic
  equations in nondivergence form.
\newblock {\em Communications in Partial differential Equations 25}, 5-6
  (2000), 821--845.

\bibitem{Freedman-75}
{\sc Freedman, D.~A.}
\newblock On tail probabilities for martingales.
\newblock {\em Annals of Probability 3\/} (1975), 100--118.

\bibitem{Guivarch-School2002}
{\sc Guivarc'h, Y.}
\newblock Limit theorems for random walks and products of random matrices.
\newblock In {\em CIMPA-TIFR School on Probability Measures on Groups: Recent
  Directions and Trends\/} (2002), M.~TIFR, Ed.

\bibitem{Hall1980}
{\sc Hall, P., and Heyde, C.~C.}
\newblock {\em Martingale limit theory and its application}.
\newblock Academic Press, 1980.

\bibitem{Lepri-Livi-Politi-2}
{\sc Lepri, S., Livi, R., and Politi, A.}
\newblock Anomalous heat conduction.
\newblock In {\em Anomalous Transport: Foundations and Applications},
  R.~Klages, G.~Radons, and I.~M. Sokolov, Eds. Wiley-VCH Verlag, Weinheim,
  2008, ch.~10.

\bibitem{Matsuda-Ishii-1970}
{\sc Matsuda, H., and Ishii, K.}
\newblock Localization of normal modes and energy transport in the disordered
  harmonic chain.
\newblock {\em Supplement of the Progress of theoretical physics 45\/} (1970),
  56--86.

\bibitem{Mustapha-2006}
{\sc Mustapha, S.}
\newblock Gaussian estimates for spacially inhomogeneous random walks on $
  z^d$.
\newblock {\em Annals of Probability 34}, 1 (2006), 264--283.

\bibitem{O'Connor-75}
{\sc O'Connor, A.~J.}
\newblock A central limit theorem for the disordered harmonic chain.
\newblock {\em Comm. Math. Phys. 45}, 1 (1975), 63--77.

\bibitem{Peierls-1929}
{\sc Peierls, R.~E.}
\newblock Zur kinetischen theorie der w{\"a}rmeleitung in kristallen.
\newblock {\em Annalen der Physik 395}, 8 (1929), 1055--1101.

\bibitem{Peierls-book-1955}
{\sc Peierls, R.~E.}
\newblock {\em Quantum Theory of Solids}.
\newblock Oxford University Press, London, 1955.

\bibitem{Raugi-1997}
{\sc Raugi, A.}
\newblock Th\'eor\`eme ergodique multiplicatif. produits de matrices
  al\'eatoires ind\'ependantes.
\newblock {\em Publ. Inst. Rech. Math. Rennes\/} (Nov 1997), 1--43.

\bibitem{Rieder-Lebowitz-Lieb-1967}
{\sc Rieder, Z., Lebowitz, J.~L., and E., L.}
\newblock Properties of harmonic crystal in a stationary nonequilibrium state.
\newblock {\em Journal of Mathematical Physics 8}, 5 (May 1967), 1073--1078.

\bibitem{Rubin-Greer-71}
{\sc {Rubin}, R.~J., and {Greer}, W.~L.}
\newblock {Abnormal Lattice Thermal Conductivity of a One-Dimensional, Harmonic
  Isotopically Disordered Crystal}.
\newblock {\em Journal of Mathematical Physics 12\/} (Aug. 1971), 1686--1701.

\bibitem{Verheggen-1979}
{\sc Verheggen, T.}
\newblock Transmission coefficient and heat conduction of a harmonic chain with
  random masses: Asymptotic estimates on products of random matrices.
\newblock {\em Commun. Math. Phys 68}, 3 (Jan 1979), 69--82.

\end{thebibliography}
\bibliographystyle{acm}

\end{document}